\documentclass[twocolumn,twocolappendix,tighten]{aastex701}
\shorttitle{MBH formation in metal-enriched galaxies}
\shortauthors{Chon et al.}
\submitjournal{ApJ}

\graphicspath{{./}{figures/}}

\newcommand{\K}{\rm K}

\usepackage{amsmath,amssymb,amsfonts}%
\usepackage{amsthm}%

\begin{document}


\title{Formation of Heavy Seed Black Holes and Little Red Dots-like Compact Clusters in Metal-enriched Star-forming Regions}

\correspondingauthor{Sunmyon Chon}

\author[orcid=0000-0002-2912-3923,gname='Sunmyon',sname='Chon']{Sunmyon Chon}
\affiliation{Max-Planck-Institut f\"ur Astrophysik, Karl-Schwarzschild-Str. 1, Garching, Garching bei M\"unchen D-85741, Germany}
\affiliation{Astronomical Institute, Graduate School of Science, Tohoku University, Aoba, Sendai 980-8578, Japan}
\email[show]{sunmyon@MPA-Garching.MPG.DE}

\author[orcid=0000-0002-4317-767X,gname='Shingo',sname='Hirano']{Shingo Hirano}
\affiliation{Department of Applied Physics, Faculty of Engineering, Kanagawa University, Kanagawa 221-0802, Japan}
\affiliation{Department of Astronomy, School of Science, University of Tokyo, Tokyo 113-0033, Japan}
\email{shingo-hirano@kanagawa-u.ac.jp}

\author[orcid=0000-0002-4848-5508,gname='Ke-Jung',sname='Chen']{Ke-Jung Chen}
\affiliation{Academia Sinica, Institute of Astronomy and Astrophysics, Taipei 106319, Taiwan, R.O.C.}
\affiliation{Heidelberger Institut für Theoretische Studien, Schloss-Wolfsbrunnenweg 35, 69118 Heidelberg, Germany}
\email{kjchen@asiaa.sinica.edu.tw}

\author[orcid=0000-0001-5976-4599,gname='Volker',sname='Springel']{Volker Springel}
\affiliation{Max-Planck-Institut f\"ur Astrophysik, Karl-Schwarzschild-Str. 1, Garching, Garching bei M\"unchen D-85741, Germany}
\email{vspringel@MPA-Garching.MPG.DE}



\begin{abstract}
The origin of supermassive black holes (SMBHs) in the early Universe remains uncertain.
While the classical direct-collapse scenario requires pristine gas exposed to intense Lyman-Werner (LW) radiation, recent JWST observations suggest that early growing BHs may reside in compact, metal-enriched environments.
We investigate early BH formation using a cosmological radiation-hydrodynamic simulation that self-consistently follows the formation and evolution of rapidly accreting supermassive stars (SMSs).
We find that intense LW radiation develops in clustered star-forming regions; by the time the first heavy seed forms, the mass-weighted internal intensity has reached $J_{21}\sim1000$.
Most seeds form in weakly metal-enriched gas with [Z/H]$\sim-3$ to $-2$, because the strongest LW fields occur in regions already enriched by previous stellar feedback.
All halos above $10^9M_\odot$ in our zoom-in region host a BH more massive than $10^5M_\odot$.
We construct stellar spectra using MIST/PHOENIX libraries for normal stars and $6000~\mathrm{K}$ blackbody spectra for bloated SMSs.
Five of eight massive clusters with stellar masses larger than $10^6~M_\odot$ exhibit intrinsically V-shaped spectra similar to those of Little Red Dots (LRDs).
In the most prominent cases, luminous SMSs dominate the red optical emission, although Balmer breaks in normal stellar populations also contribute.
We also identify off-center LRD-like systems in our simulation resembling those recently reported in strongly lensed observations.
Our results suggest that heavy-seed formation, compact-cluster formation, and LRD-like continuum emission are linked outcomes of clustered, weakly metal-enriched, strongly irradiated environments.
\end{abstract}



\section{Introduction} \label{sec:intro}
Supermassive black holes (SMBHs) are ubiquitous in the Universe and play a major role in shaping the evolution of stars and galaxies. Most mature galaxies with stellar masses above $10^{10}\,M_\odot$ host an SMBH at their centers. The tight correlations between the masses of central SMBHs and the properties of their host galaxies suggest that SMBHs and galaxies co-evolve \citep{Kormendy+2013}. Indeed, feedback from accreting SMBHs is considered essential for reproducing the observed galaxy luminosity function. Despite their fundamental importance, however, the origin of SMBHs remains largely unknown.

A key clue to understanding SMBH formation is to observe rapidly growing BHs in their early evolutionary phases. Luminous quasars (QSOs) have already been discovered at $z \gtrsim 7$ \citep[e.g.,][]{Mortlock+2011, Wu+2015, Banados+2018, Wang+2021, Bogdan+2024}. This indicates that SMBHs with masses of $10^9$--$10^{10}\,M_\odot$ had already formed within the first several hundred million years of cosmic history. One promising pathway for forming such early SMBHs is the ``Direct Collapse'' (DC) model, in which supermassive stars (SMSs) with masses of $\sim 10^5\,M_\odot$ directly collapse into BHs of comparable mass \citep[e.g.,][]{Bromm+2003, Latif+2013, Regan+2016, Chon+2018}. The DC model requires low metallicity and strong Lyman-Werner (LW) radiation, which dissociates H$_2$ molecules and suppresses cooling and subsequent fragmentation. Once such heavy seed BHs form, sustained Eddington accretion can allow them to grow into the SMBHs observed in high-redshift QSOs.

Recent JWST observations have revealed a new population of possible early accreting BH systems, known as ``Little Red Dots'' \citep[LRDs; e.g.,][]{Kocevski+2023, Ubler+2023, DeGraf+2024, Matthee+2024, Taylor+2025}. LRDs are spatially compact sources with red colors, often characterized by V-shaped continuum spectra. Spectroscopic observations of LRDs have revealed broad H$\alpha$ emission lines, suggesting that they host accreting BHs and may represent low-luminosity active galactic nuclei (AGNs) in the early Universe \citep{Hviding+2026}. In addition, the red colors and strong Balmer breaks seen in some LRDs suggest that these systems may be embedded in dense gas environments \citep[e.g.,][]{Naidu+2025,Chang+2025}.

While the nature of LRDs is still debated, they may provide a unique view of seed BHs during their early growth phase. Observations suggest that some early growing BHs may be embedded in extremely dense gas with gas densities above $10^{10}\,\mathrm{cm^{-3}}$. The time variability of LRDs also implies that the disk or density structures around these accreting BHs may differ from those in ordinary AGNs or QSOs \citep{Kokubo&Harikane2025, Zhang+2025}. These findings have motivated several physical models for LRDs, including scenarios in which BHs accrete at super-Eddington rates \citep{Inayoshi_dense_gas+2025}, as well as BH-star models, in which a BH is surrounded by dense gas originating from the envelope of its progenitor SMS \citep[e.g.,][]{Naidu+2025}.

LRDs are often found in association with nearby star-forming galaxies. This raises the possibility that they are linked to DCBH formation, in which strong LW radiation from neighboring galaxies triggers the formation of 
heavy seed BHs. \citet{Naidu+2025} found an LRD located at a projected distance of $60\,\mathrm{kpc}$ from a star-forming galaxy. More statistically, \citet{Baggen+2026} reported that $40$--$80\%$ of LRDs have nearby ``blue dots'', depending on the LRD luminosity. These blue dots may be star-forming galaxies located at separations of $0.5$--$5\,\mathrm{kpc}$ from the LRDs. The estimated LW intensity at the positions of the LRDs can reach $J_{21} \gtrsim 1000$, which is strong enough to trigger DCBH formation \citep{Chon+2016}. 
Recent observations of strongly lensed systems further reveal associations on much smaller scales. \citet{Yanagisawa+2026} found clustered LRDs located at projected distances of only $\sim 70\,\mathrm{pc}$ from the center of a star-forming galaxy, a configuration that would remain unresolved in typical observations. Such close proximity to star-forming galaxies may lead to significant metal enrichment, posing a challenge to the classical DC scenario.

Numerical simulations have shown that DCBHs can form in strongly irradiated atomic-cooling halos. While the number density estimated from the DC model has a large uncertainty \citep{Habouzit+2016}, the cosmological simulation by \citet{Chon+2016} shows the DCBH density to be $\sim 10^{-4}\,\mathrm{Mpc^{-3}}$ , which is comparable to the observed number density of LRDs \citep{Harikane+2023, Matthee+2024, Taylor+2025}. Cosmological radiation-hydrodynamic simulations by \citet{Chon+2026} have shown that DCBHs are expected to be surrounded by dense gas with the gas densities of $10^{10}$--$10^{11}\,\mathrm{cm^{-3}}$. This dense gas efficiently feeds the DCBHs and allows them to grow at super-Eddington rates, rapidly increasing the BH mass to $10^7\,M_\odot$ within the first $0.1\,\mathrm{Myr}$. 

At the small separations observed for some LRDs, metal enrichment from nearby star formation may inhibit classical DCBH formation. Heavy seed BHs  may nevertheless form through the ``super-competitive accretion'' (SCA) scenario, which can operate in metal-enriched environments. \citet{Chon+2025} conducted radiation-hydrodynamic simulations resolving protostellar scales of $10$--$100\,\mathrm{au}$. They found that massive BH formation can occur even at a metallicity of $10^{-3}\,Z_\odot$. Although cooling by metals and dust grains induces vigorous fragmentation, massive gas inflows and stellar mergers allow central stars to grow to $10^4$--$10^5\,M_\odot$. \citet{Chiaki+2023} estimated the number density of heavy seed BHs produced by the SCA scenario using a semi-analytic model and found it to be about two orders of magnitude higher than that expected from the classical DC model. However, it remains unclear whether the conditions required for the SCA scenario naturally emerge in the environments of forming galaxies in cosmological hydrodynamic simulations.

In this paper, we report the formation of heavy seed BHs in the outskirts of galaxies, consistent with the SCA scenario. Massive gas inflows lead to the formation of supermassive stars and heavy BH seeds with masses of $10^5$--$10^6\,M_\odot$. 
Bloated SMSs in these systems can reproduce the compact morphology and V-shaped intrinsic continuum of the observed off-centered lensed LRD sources \citep{Yanagisawa+2026}.

\section{Methodology} \label{sec:method}
We performed zoom-in radiation-hydrodynamic simulations using the open-source code {\tt AREPO} \citep{Springel2010} with several extensions. The cosmological initial conditions were generated at $z=99$ using {\tt NGenIC}, which is  built in {\tt Gadget4} \citep{Springel+2021}. To identify the target region for the zoom-in simulation, we first performed a base N-body simulation in a cubic cosmological volume with a side length of $128\,h^{-1}\mathrm{Mpc}$. The base simulation contained $1024^3$ particles and was evolved to $z=15$. We identified the most massive halo of $10^{10}~h^{-1}M_\odot$ at $z=15$, traced its particles back to the initial conditions, and defined their center of mass as the center of the zoom-in region. We then defined a cubic zoom-in region with a side length of $1~h^{-1}\mathrm{Mpc}$ and refined its initial Lagrangian volume to an effective resolution of $32,768^3$. Within the zoom-in region, the dark matter particle mass and the initial gas-cell mass were $4300$ and $800~h^{-1}M_\odot$, respectively. Outside the zoom-in region, the initial resolution was reduced to an effective resolution of $256^3$. The volume of the parent simulation is large enough to contain UV-luminous galaxies at $z\gtrsim12$ \citep{Harikane+2023}.

During the simulation, cells were refined whenever the local Jeans length was resolved by fewer than 32 cell radii. We largely followed the numerical methods described in \citet{Chon+2026} and briefly summarize the numerical setup below.

\subsection{Chemistry}
We followed a non-equilibrium primordial chemistry network including e$^{-}$, H, H$^{+}$, H$_2$, H$^{-}$, D, D$^{+}$, and HD, together with the associated cooling processes \citep{Matsukoba+2021, Chon+2021}. In addition, we included photodissociation of H$_2$ by Lyman--Werner (LW) radiation and photodetachment of H$^{-}$ by infrared photons.

After the onset of metal enrichment, we also included metal and dust chemistry and cooling. Specifically, we considered fine-structure cooling by C~{\sc ii} and O~{\sc i}, thermal emission from dust grains, and H$_2$ formation on dust-grain surfaces. We assumed that the elemental abundance pattern, dust-to-metal mass ratio, grain composition, and grain-size distribution were identical to those in the solar neighborhood. We further assumed that all gas-phase carbon and oxygen were present as C~{\sc ii} and O~{\sc i}, respectively. This approximation reproduces the thermal evolution of collapsing clouds over a wide range of metallicities, from $10^{-6}\,Z_\odot$ to $Z_\odot$ \citep{Chon+2024}. The dust temperature was determined by balancing dust heating and cooling, including radiative heating by the CMB and nearby stars.

\subsection{Star formation and evolution}
\subsubsection{star formation}
We created sink particles when gas cells simultaneously satisfied the following three conditions:
1) the gas density exceeded the threshold density, $n_\text{th}=2\times10^5\,\mathrm{cm^{-3}}$;
2) the cell was located at a local minimum of the gravitational potential; and
3) the cell was gravitationally bound.
We set the sink radius to four times the local Jeans length, resulting in sink radii of $1$--$10\,\mathrm{pc}$. Because the density threshold for sink formation was fixed, the variation in sink radius primarily reflected differences in the gas temperature at sink formation. We assumed that each sink particle represented a single protostar.

We modeled the properties of accreting protostars in two evolutionary phases:
1) a normal stellar-evolution phase, during which the star evolved from the pre-main-sequence stage, underwent Kelvin--Helmholtz contraction, and reached the main sequence; and
2) a supergiant phase, in which an accretion rate above $\dot{M}_\text{crit}=0.02\,M_\odot\,\mathrm{yr^{-1}}$ maintained an inflated stellar envelope \citep{Hosokawa+2012, Hosokawa+2013}.
We assigned the stellar radius, luminosity, and effective temperature separately in the two phases. During the normal stellar-evolution phase, these properties were determined by extrapolating the tabulated results of \citet{Hosokawa+2009}, which provide stellar properties as functions of stellar mass and accretion rate. During the supergiant phase, we assumed an effective temperature of $6000\,\mathrm{K}$ and the Eddington luminosity.

We determined the stellar structure according to the following prescription. Once the accretion rate exceeded $\dot{M}_\text{crit}$, the star entered the supergiant phase. After the accretion rate fell below this threshold, the stellar radius gradually approached its main-sequence value over the surface Kelvin--Helmholtz timescale. Following \citet{Sakurai+2015}, we adopted
\begin{align}
t_\text{KH,surface}=10t_\text{KH}
\equiv \frac{10GM_*^2}{R_*L_*}.
\end{align}

We estimated the final stellar mass using the relation between the mass inflow rate on parsec scales and the final stellar mass, motivated by high-resolution radiation-hydrodynamic simulations \citep{Hirano+2015, Toyouchi+2023}:
\begin{align}
M_* &= 250\,M_\odot
\left(
\frac{\dot{M}_*}{2.8\times10^{-3}\,M_\odot\,\mathrm{yr^{-1}}}
\right)^{0.7},
\end{align}
where $\dot{M}_*$ was estimated from the mass accretion rate during the first $10^4\,\mathrm{yr}$.
Although this relation was originally derived for primordial stars, we also apply it to the finite-metallicity cases considered here. The stellar metallicity reaches only up to [Z/H]$=-2$, where radiation pressure on dust remains subdominant for mass accretion \citep{Fukushima+2020}. We therefore expect the relation between the parsec-scale accretion rate and the final stellar mass to remain approximately applicable in these metal-poor environments.

We refer to star particles with masses larger than $10^4~M_\odot$ as supermassive stars (SMSs). At [Z/H]$\lesssim-3$, this interpretation is supported by previous high-resolution simulations showing that rapidly accreting central stars can grow to SMS masses under comparable inflow conditions \citep{Chon+2020, Chon+2025}. At higher metallicities, particularly around [Z/H]$\sim-2$, unresolved fragmentation may redistribute the mass of a massive star particle among multiple stars, producing a compact, highly top-heavy stellar system instead. Nevertheless, we retain the term SMS as an operational classification throughout this paper and discuss this uncertainty in Section~\ref{sec:discussion1}.

We assumed that SMSs collapsed directly into BHs and classified the resulting remnants as heavy-seed BHs if the final SMS masses exceeded $10^4\,M_\odot$. We did not include subsequent gas accretion onto these BHs.

\subsubsection{metal enrichment}
We adopt stellar lifetimes as a function of the stellar mass from the zero-metallicity stellar evolution models of \citet{Schaerer2002}.
We determined the fate of each star after the stellar lifetime according to its final mass, following \citet{Heger+2002}. Stars with $8\,M_\odot\leq M_*\leq40\,M_\odot$ were assumed to explode as core-collapse supernovae (SNe) with an explosion energy of $10^{51}\,\mathrm{erg}$. Stars with $140\,M_\odot\leq M_*\leq260\,M_\odot$ were assumed to undergo pair-instability supernovae (PISNe) with an explosion energy of $10^{53}\,\mathrm{erg}$. Stars with $40\,M_\odot<M_*<140\,M_\odot$ or $M_*>260\,M_\odot$ were assumed to collapse directly into BHs without SN explosions.

The ejected metal mass was adopted from \citet{Nomoto+2013}:
\begin{align}
\frac{M_\text{eje}}{M_\odot}
&= \frac{M_*}{M_\odot}-80,
& (M_*>140~M_\odot),\\
&=0.343\left(\frac{M_*}{M_\odot}-8\right)+0.1,
& (M_*<40~M_\odot).
\end{align}
The injection radius, $R_\text{inj}$, was defined to enclose 20 gas cells, over which the metal ejecta were distributed. We injected the SN energy thermally when the Sedov--Taylor radius was resolved \citep{Magg+2022}, where
\begin{align}
R_\text{Sedov}
=24.0~\mathrm{pc}
\left(\frac{n}{1~\mathrm{cm^{-3}}}\right)^{-0.42}
\left(\frac{E_\text{SN}}{10^{51}~\mathrm{erg}}\right)^{0.29},
\end{align}
and $n$ and $E_\text{SN}$ are the gas number density and SN explosion energy, respectively. If $R_\text{inj}<R_\text{Sedov}$, we injected thermal energy into the cells within the injection region; otherwise, we injected momentum.

\subsection{Radiation transfer}
\label{sec:rt}

We followed the propagation of photons using the tree-based ray-tracing method {\tt START} \citep{Hasegawa+2010}. This method is an extension of {\tt RSPH}, in which the optical depth is evaluated by summing local contributions along each ray \citep{Susa2006, Chon+2017}. We constructed an octree containing the radiation sources. When the opening angle of a source node was smaller than $\Theta_\text{crit}\equiv0.7$, the sources within that node were combined. We summed their luminosities and calculated the luminosity-weighted mean position and effective temperature of the combined source. This procedure reduces the cost of ray tracing, with the computational cost scaling only as $\propto\ln N_\text{source}$, where $N_\text{source}$ is the number of sources.

We calculated the optical depth to ionizing radiation, $\tau_\text{ion}$, the optical depth due to scattering by gas, $\tau_\text{scatter}$, the optical depth due to scattering and absorption by dust grains, $\tau_\text{dust}$, and the column densities of H$_2$, $N_{\mathrm{H}_2}$. For ionizing radiation, we adopted the photoionization cross-section
\begin{align}
\sigma_\nu
=6.3\times10^{-18}
\left(\frac{\nu}{\nu_0}\right)^{-3}\,\mathrm{cm^2},
\end{align}
where $h\nu_0=13.6\,\mathrm{eV}$. We evaluated $\tau_\text{scatter}$ and $\tau_\text{dust}$ at $11.2\,\mathrm{eV}$ and neglected their frequency dependence. The spectral energy fluxes of EUV and LW radiation from a single source at a distance $r$ were calculated as
\begin{align}
F_{\nu,\text{EUV}}
&=\frac{L_\nu}{4\pi r^2}
\exp\left[-\tau_\text{ion}(\nu)-\tau_\text{scatter}
-\tau_\text{dust}\right],\\
F_{\nu,\text{LW}}
&=\frac{L_\nu}{4\pi r^2}
\exp\left[-\tau_\text{scatter}-\tau_\text{dust}\right]
f_\text{sh}(N_{\mathrm{H}_2}),
\end{align}
where the H$_2$ self-shielding factor is given by
\begin{align}
f_\text{sh}(N_{\mathrm{H}_2})
=\min\left[
1,
\left(\frac{N_{\mathrm{H}_2}}{N_0}\right)^{-3/4}
\right],
\end{align}
with $N_0=10^{14}\,\mathrm{cm^{-2}}$ \citep{Draine+1996}. We assumed that the spectrum of each individual source was described by a blackbody with an effective temperature $T_\text{eff}$.

We calculated the photodetachment rate of H$^{-}$ in the optically thin limit. We accounted for geometric dilution of the radiation field by assuming $T_\text{rad}^4\propto r^{-2}$ which is used to estimate the dust-irradiation effect \citep{Offner+2010}.

\subsection{Cluster identification and spectra}
We identified star clusters using a friends-of-friends algorithm with a linking length of $1$ comoving $h^{-1}\,\mathrm{kpc}$ and a minimum membership of 20 star particles. At the final snapshot, we identified 34 clusters, eight of which had total stellar masses larger than $10^6~M_\odot$. We label the clusters with IDs from 0 to 33 in order of decreasing stellar mass.

\subsubsection{Reconstruction of stellar IMF}\label{sec:reconstruction}
We constructed the spectra separately for normal stars and SMSs, adopting $10^4\,M_\odot$ as the threshold SMS mass. For each star particle, we divided its mass into a normal-stellar component that follows Chabriere IMF \citep{Chabrier2003} and a log-flat high-mass component according to the adopted value $f_{\rm Ch}=0.1$ (see Section~\ref{sec:IMF}). The normal-stellar component was evaluated through Monte Carlo sampling with mass-conserving weights. The sampled stars should therefore be regarded as representative samples used to integrate the stellar population, rather than as a strictly discrete realization of all low-mass stars. Since the low-mass population contains a large number of stars, its integrated spectral contribution is statistically well sampled and is only weakly affected by Monte Carlo noise.

For the log-flat component, we explicitly sampled individual stars with unit weight until the remaining mass fell below the SMS threshold. The residual mass was assigned using the same mass-conserving weighted prescription as for the normal component. Consequently, SMSs were treated as discrete Monte Carlo objects, whereas the normal stellar component was represented by a weighted Monte Carlo integration of the IMF. This distinction is important for the spectral analysis because SMSs are rare but can dominate the emergent spectrum; we therefore sample them explicitly as individual objects rather than averaging their contribution over the IMF. 

\subsubsection{Cluster spectra}
For normal stars, we used a precomputed stellar library constructed from the MESA Isochrones and Stellar Tracks (MIST) models \citep{Dotter2016, Choi+2016} and the PHOENIX stellar-atmosphere library. The MIST models provide the effective temperature, surface gravity, and bolometric luminosity as functions of initial stellar mass, age, and metallicity, while the PHOENIX models provide the corresponding spectral shapes. For each surviving Monte Carlo star, we interpolated the library in stellar mass, age, and metallicity and added its contribution to the cluster spectrum. The MIST-based library extends up to $300\,M_\odot$. For normal stars more massive than $300\,M_\odot$, we adopted the spectral shape of the $300\,M_\odot$ model and renormalized it to the Eddington luminosity corresponding to the actual stellar mass.

Stars more massive than $10^4\,M_\odot$ were classified as SMSs. We modeled their spectra as blackbody emission with an effective temperature of $6000\,\mathrm{K}$ and a bolometric luminosity equal to the Eddington luminosity \citep{Hosokawa+2013, Nandal+2024}. We first calculated the intrinsic stellar spectra without dust attenuation or dust re-emission. For comparison, we subsequently applied an approximate foreground-attenuation model based on the spherically averaged dust column density and the Calzetti attenuation law \citep{Calzetti+2000}. Dust re-emission was not included.

\begin{figure*}
\centering
\includegraphics[width=0.95\textwidth]{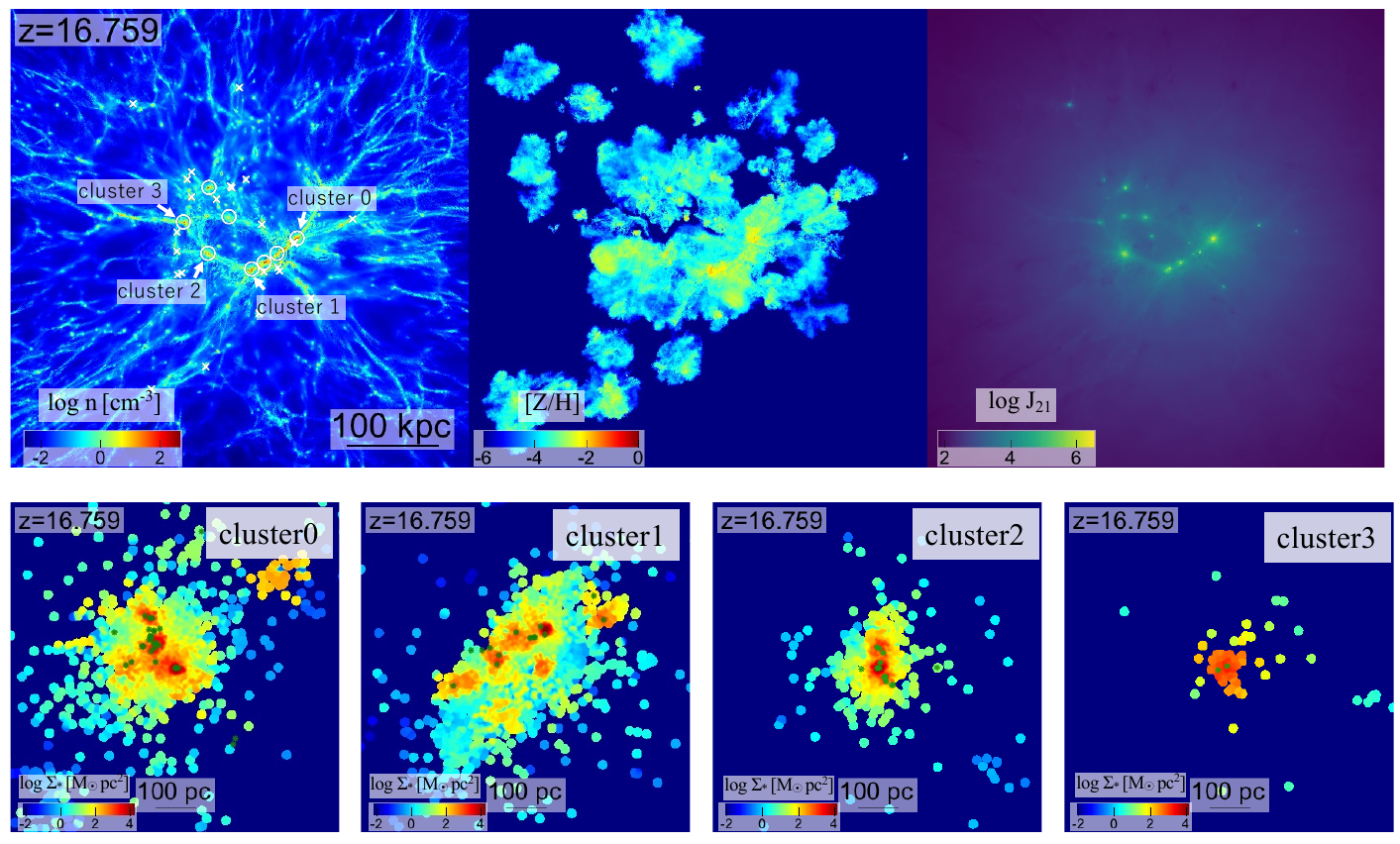}
\caption{
Top panels: Projected distributions of gas density (left), metallicity (middle), and FUV intensity $J_{21}$ (right) at the end of the simulation at $z=16.759$. White circles indicate the eight most massive star clusters with stellar masses larger than $10^6\,M_\odot$, while crosses indicate the remaining 26 clusters. The eight massive clusters have host SMSs with masses larger than $10^4~M_\odot$. The length scales are shown in comoving units.
Bottom panels: Spatial distributions of star particles within the four most massive clusters, clusters 0--3. The colors indicate the stellar surface density. Green asterisks mark the positions of surviving SMSs with masses larger than $10^4~M_\odot$. The length scales are shown in physical units.
}\label{fig:overall}
\end{figure*}

\begin{figure}[t]
\centering
\includegraphics[width=0.45\textwidth]{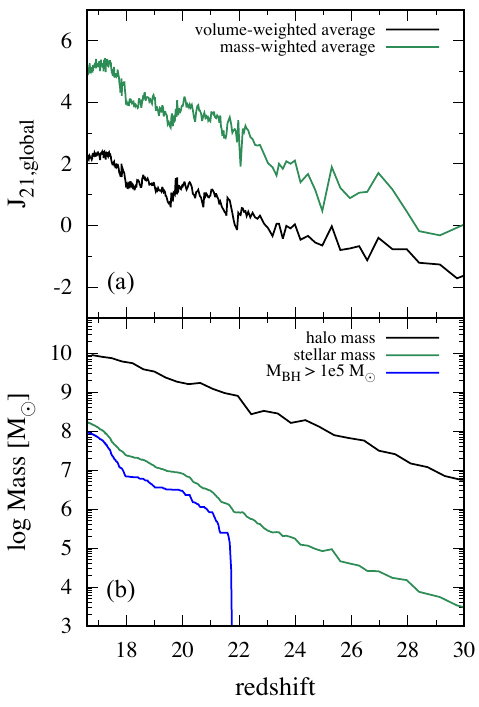}
\caption{
(a) Redshift evolution of the LW intensity, $J_{21,\mathrm{global}}$, averaged over the zoom-in region. The black and green lines show the volume-weighted and mass-weighted averages, respectively. 
(b) Redshift evolution of the mass of the most massive halo (black), stars (green), and BHs with masses larger than $10^5~M_\odot$ (blue).
}\label{fig:evolution}
\end{figure}

\begin{figure}[t]
\centering
\includegraphics[width=0.5\textwidth]{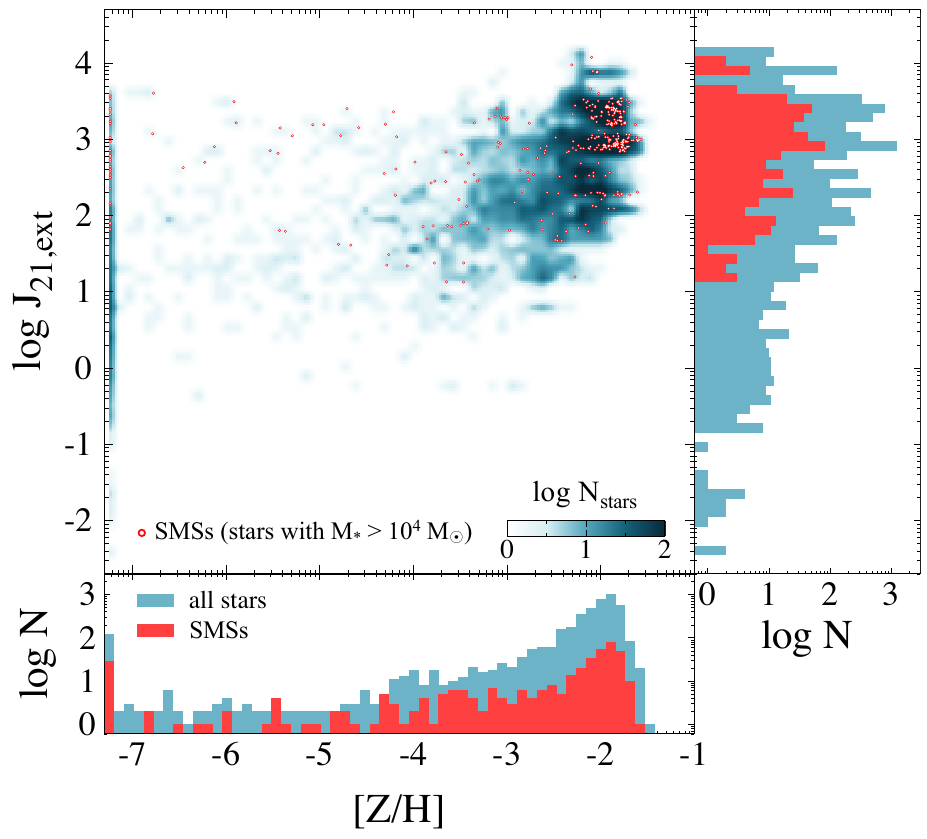}
\caption{
Distribution of the external LW intensity $J_{21,\mathrm{ext}}$ at the time of star formation and the stellar metallicity for all stars formed in the simulation. The colors indicate the number of stars in each bin. The distribution of SMSs (stars with masses larger than $10^4~M_\odot$) is overplotted by red circles. When calculating the external LW intensity, we exclude LW sources located within the same halo.
The histograms show the distributions of the external LW intensity and stellar metallicity for all stars (blue) and for stars with masses larger than $10^4~M_\odot$ (red).
}\label{fig:J21_Mstar}
\end{figure}

\begin{figure} [t]
\centering
\includegraphics[width=0.47\textwidth]{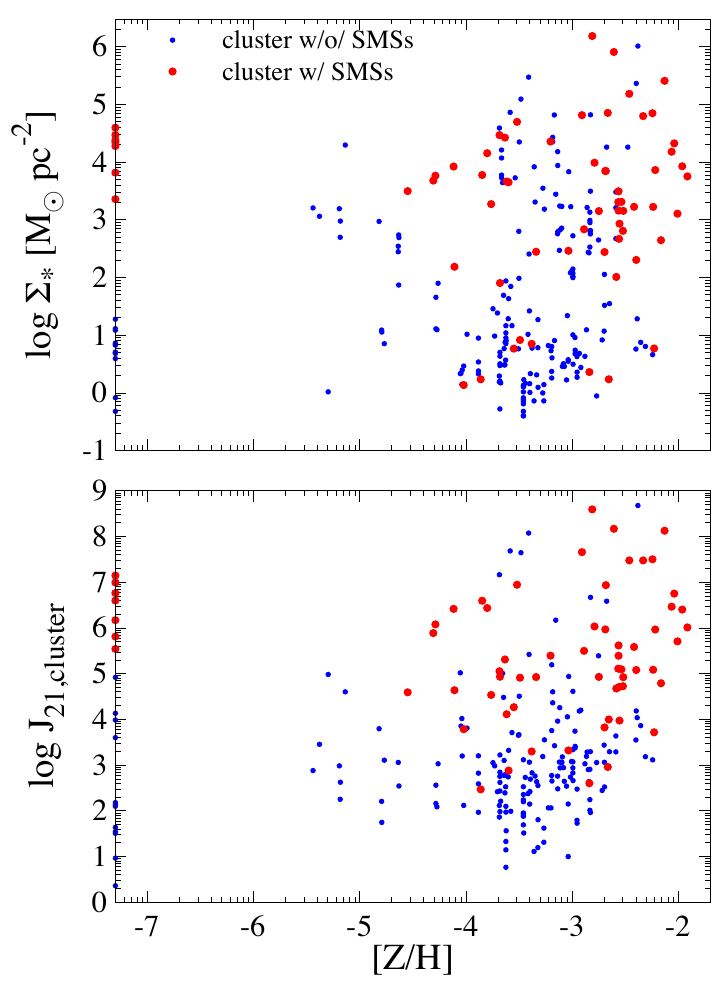}
\caption{
Scatter plots of the cluster properties. The top panel shows the stellar surface density as a function of cluster metallicity, while the bottom panel shows the volume-weighted median FUV intensity, $J_{21,\mathrm{cluster}}$, as a function of cluster metallicity. Unlike $J_{21,\mathrm{ext}}$ defined in Fig.~\ref{fig:J21_Mstar}, $J_{21,\mathrm{cluster}}$ includes the contribution from local sources within the same halo. Here, the cluster metallicity is defined as the gas-phase metallicity within the cluster, with values below $5\times10^{-7}\,Z_\odot$ set to this floor value.
Red points indicate clusters that form SMSs with masses larger than $10^4~M_\odot$ by the next snapshot, while blue points indicate clusters without subsequent SMS formation.
}\label{fig:Sigma_J21_Z_relation}
\end{figure}

\begin{figure*}[t]
\centering
\includegraphics[width=0.9\textwidth]{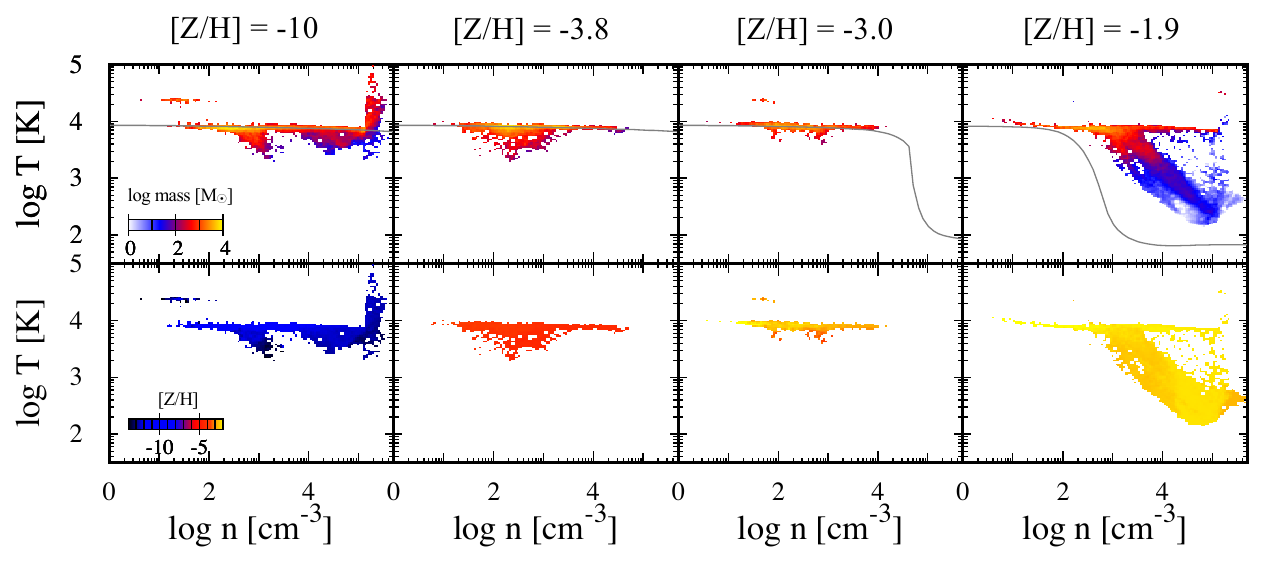}
\caption{
Phase diagrams of temperature versus gas density for four different environments that form SMSs with masses larger than $10^5~M_\odot$. The snapshots are taken immediately before SMS formation, and we select gas cells within $50~\mathrm{pc}$ of the cell that will be converted into the protostar. The colors indicate the gas mass in each cell (top panels) and the gas metallicity (bottom panels).
We select four cases according to the metallicity of the newly formed
star particle: [Z/H]$=-10$, $-3.8$, $-3.0$, and $-1.9$, from left to right. The thin grey line shows the temperature evolution predicted by the 1-zone model, in which the thermal evolution of a collapsing cloud is solved assuming collapse on the free-fall timescale.
}\label{fig:rhoT_hist}
\end{figure*}

\section{Results} \label{sec:results}
\subsection{Overall evolution}
The top panels of Fig.~\ref{fig:overall} show the projected distributions of gas density, metallicity, and FUV intensity, $J_{21}$, in the zoom-in region at $z=16.759$. By this epoch, several star-forming clusters have formed within the central $\sim 100~\mathrm{kpc}$ region. We identify eight massive clusters with stellar masses larger than $10^6~M_\odot$, which are marked by white circles. These clusters have already been enriched to metallicities [Z/H]$=-3$ to $-2$, except for one extremely low-metallicity cluster with [Z/H]=$-7$ (Cluster 5, see Section~\ref{sec:cluster_statistics}). The intense star formation also produces a strong FUV radiation field around the central star-forming regions, where the intensity reaches $J_{21}\gtrsim 10^4$.

The simulation produces compact clusters with sizes consistent with recent JWST observations \citep{Harikane+2023}. The bottom panels show the spatial distributions of stars and the stellar surface densities in clusters 0--3. Clusters 0 and 1 have extended structures spanning more than $100$ physical pc, whereas clusters 2 and 3 exhibit more compact morphologies with sizes comparable to or smaller than $100$ physical pc. Their stellar surface densities range from $10^2$ to $10^4~M_\odot\,\mathrm{pc}^{-2}$. The clusters also contain clumpy substructures whose peak surface densities reach $10^4~M_\odot\,\mathrm{pc}^{-2}$, resembling the compact star clusters observed in strongly lensed systems \citep[e.g.,][]{Adamo+2024}.

The strong FUV radiation field promotes the formation of SMSs within these clusters, which are assumed to subsequently collapse into heavy-seed BHs. Green asterisks indicate the positions of surviving SMSs with masses larger than $10^4~M_\odot$. Their formation is triggered by intense FUV radiation from nearby stars, which photodissociates H$_2$ molecules and raises the gas temperature to approximately $10^4~\mathrm{K}$, as predicted by the direct-collapse model \citep{Bromm+2003}. SMSs form not only near the centers of the clusters but also at the outskirts of the galaxy, as seen in cluster 1. In Sec.~\ref{sec:spectra}, we discuss the similarity between this configuration and the recently observed ``Red Eyes'' system, in which LRDs are located several hundred parsecs from star-forming clumps \citep{Yanagisawa+2026}.

\subsection{Formation of SMSs and heavy seed BHs}
Fig.~\ref{fig:evolution} shows (a) the time evolution of the LW intensity, $J_{21,\text{global}}$, averaged over the zoom-in region, and (b) the growth of the most massive halo, the stellar component, and the heavy seed BH population.
The most massive halo grows steadily and reaches $\sim 10^{10}~M_\odot$ by $z\simeq 17$, which contains the eight most massive clusters.
The first heavy seed BH forms at $z\sim 22$, after which the total mass in heavy seed BHs continues to increase. 
We note that we do not include subsequent gas accretion onto the seed BHs. 
Therefore, the increase in the total BH mass shown in panel (b) is entirely due to the formation of additional heavy seed BHs, rather than the growth of individual BHs.

\begin{figure}[t]
\centering
\includegraphics[width=0.45\textwidth]{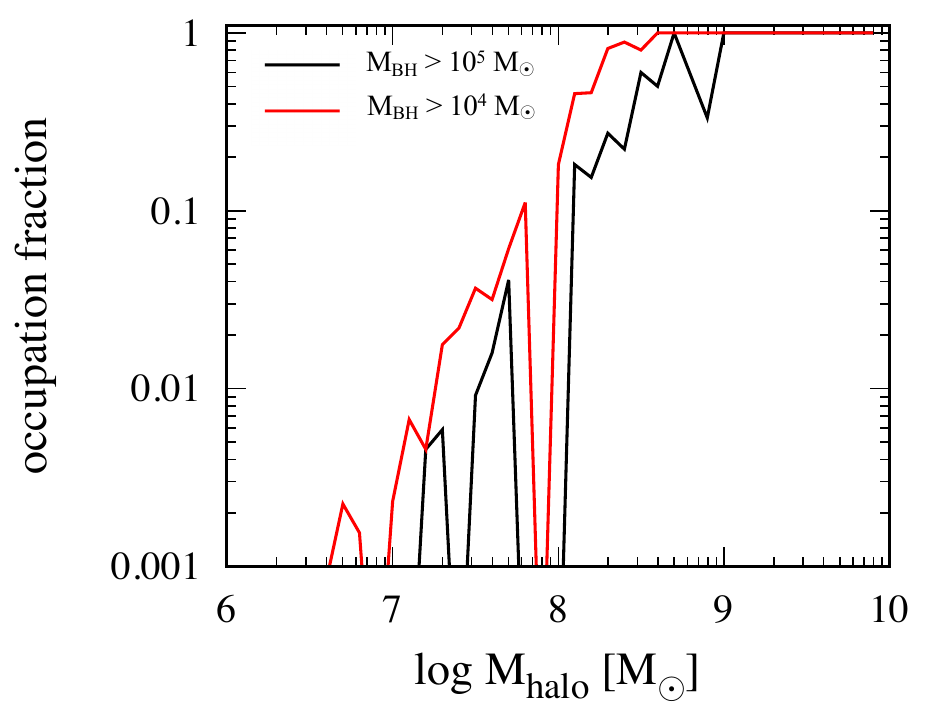}
\caption{
Halo occupation fraction of heavy seed BHs as a function of halo mass. The red and black lines show the occupation fractions of BHs with masses larger than $10^4~M_\odot$ and $10^5~M_\odot$, respectively. We compile halos from all snapshots and calculate the fraction of halos hosting at least one BH above each mass threshold.
}\label{fig:BH_occupation_fraction}
\end{figure}

Panel (a) shows that the LW radiation field has already become strong by the time the first heavy-seed BH forms at $z \sim 22$. The mass-weighted average of $J_{21}$ is particularly relevant for BH formation because it traces the radiation field experienced by dense gas around star-forming regions, rather than that in the volume-filling diffuse medium. Around the epoch of the first heavy-seed BH formation, the mass-weighted average reaches $J_{21,\text{global}}\sim 10^3$, comparable to the critical LW intensity often required for direct-collapse BH formation \citep{Shang+2010, Sugimura+2014, Latif+2014}. This indicates that the gas relevant to SMS and heavy-seed formation is exposed to an intense LW radiation field, which suppresses H$_2$ cooling through photodissociation. Although this global mass-weighted intensity is not directly equivalent to the local irradiating intensity used to define $J_{21,\rm crit}$, the two quantities reach comparable orders of magnitude.

In contrast, the volume-weighted average remains much lower, at $J_{21,\mathrm{global}}\sim 1$ when the first heavy-seed BH forms at $z\sim 22$. This is because most of the zoom-in volume is occupied by low-density gas far from the main star-forming regions. The large difference between the mass-weighted and volume-weighted averages therefore indicates that the LW radiation field is highly inhomogeneous. Strong LW irradiation is localized around dense, clustered star-forming structures, whereas most of the volume experiences only a weak background radiation field.

Fig.~\ref{fig:J21_Mstar} shows the relation between the stellar metallicity, [Z/H], and the external LW intensity, $J_{21,\text{ext}}$, for all star particles formed in our simulation. We overplot the distribution of SMSs. The stellar metallicity and LW intensity are positively correlated because star formation increases both the metallicity and the FUV intensity within clusters.

The LW intensity at the formation sites of SMSs is biased toward higher values. This is also evident in the one-dimensional histogram of $J_{21,\text{ext}}$, where the blue histogram shows the distribution for all stars, while the red histogram shows that for SMSs. SMSs form only in strongly irradiated environments, where $J_{21,\text{ext}}\gtrsim 10$. This result supports the picture that strong LW radiation plays a key role in suppressing H$_2$ cooling and enabling the formation of very massive stars.

The external LW intensity associated with heavy-seed formation in our simulation is somewhat lower than the commonly quoted critical range, $J_{21,\rm crit}=100$--$1000$ \citep{Shang+2010, Sugimura+2014, Latif+2013}. This difference is partly due to our definition of the external LW intensity. When estimating $J_{21,\text{ext}}$, we exclude contributions from stars located within the same halo as the collapsing object. This definition is intended to remove short-range contributions from stars formed through local fragmentation, which can otherwise dominate the LW intensity because of the $J_{21,\text{ext}}\propto r^{-2}$ scaling, where $r$ is the distance from the LW source.

Such local LW sources do not necessarily promote DCBH formation. They are usually short-lived and may only delay cloud collapse without substantially altering the large-scale accretion flow \citep{Sullivan+2025}. However, this procedure is conservative and may remove part of the physically relevant radiation field. Some stars within the same halo may still act as genuine external LW sources for the collapsing cloud, particularly when they are spatially separated from the SMS-forming clump. Therefore, $J_{21,\text{ext}}$ shown in Fig.~\ref{fig:J21_Mstar} should be regarded as a lower-limit estimate of the LW intensity relevant to suppressing H$_2$ cooling. Including all nearby sources would shift some systems toward higher $J_{21}$ and reduce the apparent offset from the conventional $J_{21,\rm crit}$ range.

Fig.~\ref{fig:J21_Mstar} also shows that many SMSs form from gas that has already been metal-enriched. The bottom histogram of Fig.~\ref{fig:J21_Mstar} shows the metallicity distributions of all stars and SMSs. The metallicity distribution of all stars is bimodal. The first peak occurs at very low metallicity and corresponds to Pop~III stars, whereas the second peak occurs around [Z/H]$=-2$ and corresponds to stars formed in the central clustered regions shown in Fig.~\ref{fig:overall}. The latter peak indicates that SN feedback rapidly enriches these regions to metallicities of [Z/H]=$-3$ to $-2$.

The SMS population broadly follows the metallicity distribution of all stars, although the number of SMSs increases above metallicities of [Z/H]=$-3$ to $-2$. This increase partly reflects the fact that regions exposed to the strongest LW radiation have typically already been enriched by previous episodes of star formation. Strong LW irradiation and metal enrichment are therefore spatially correlated in our simulation.

\subsubsection{Cluster condition to form SMSs}
We quantify the conditions under which clusters form SMSs using a sample of 260 clusters identified across 41 snapshots. We classify a cluster as SMS-forming if it forms a new SMS by the next snapshot, yielding 65 SMS-forming clusters in total. This definition avoids a spurious positive correlation between SMS formation and $J_{21,\mathrm{cluster}}$, because pre-existing SMSs themselves produce intense LW radiation. The snapshot interval is typically $6\,\mathrm{Myr}$, comparable to the free-fall time of gas at densities of $10$--$100\,\mathrm{cm^{-3}}$, and therefore probes whether a cluster proceeds to SMS formation on the characteristic collapse timescale.

Fig.~\ref{fig:Sigma_J21_Z_relation} shows the stellar surface density as a function of cluster metallicity in the top panel and $J_{21,\mathrm{cluster}}$, defined as the median $J_{21}$ within each cluster, as a function of cluster metallicity in the bottom panel. Blue points indicate clusters without subsequent SMS formation, whereas red points indicate SMS-forming clusters. Both the stellar surface density and $J_{21,\mathrm{cluster}}$ correlate strongly with cluster metallicity. SMS formation occurs only in clouds that contain a sufficiently large mass of Jeans-unstable gas and remain gravitationally unstable for a sufficiently long time.

The top panel shows that SMS-forming clusters are biased toward more compact systems. Of the 65 SMS-forming clusters, 56 have mean stellar surface densities larger than $100~M_\odot~\mathrm{pc^{-2}}$. The bottom panel shows that SMS formation preferentially occurs in clusters with higher $J_{21,\text{cluster}}$. Of the 65 SMS-forming clusters, 61 have median intensities of $J_{21,\text{cluster}}>1000$. This is consistent with the critical LW intensity commonly required for heavy-seed BH formation. This indicates that the SMS formation is more strongly related to $J_{21,\text{cluster}}$ than the stellar surface density, suggesting that the local FUV radiation field is a more direct driver of SMS formation. Although clusters with higher stellar surface densities are more likely to form SMSs, the stellar surface density reflects the integrated star formation history and does not necessarily trace the recent star formation rate or the local $J_{21}$ field.

\subsubsection{Metallicity dependence of SMS formation}
\label{sec:individual_SMS_cases}
Fig.~\ref{fig:rhoT_hist} shows the gas density--temperature phase diagrams of SMS progenitor clouds for four cases selected according to the metallicity of the newly formed star particle, [Z/H]$=-10$, $-3.8$, $-3.0$, and $-1.9$. The top panels show the gas mass distribution, while the bottom panels show the gas metallicity in the density--temperature plane. For all the metallicity cases, the temperature initially evolves along the atomic cooling path, keeping the temperature of $10^4~$K below $n\lesssim 10^2$--$10^3~\mathrm{cm^{-3}}$. The later evolution above that density depends on the cloud metallicity.

At metallicities below [Z/H]$\simeq -3$, the gas evolves nearly isothermally at a temperature of approximately $10^4~\mathrm{K}$ because fine-structure line cooling is too weak to reduce the temperature over this density range. This thermal evolution is consistent with that predicted by the one-zone model, shown by the thin grey lines. In this model, the thermal evolution of a collapsing cloud is calculated assuming collapse on the free-fall timescale. At [Z/H]$\simeq -2$, fine-structure line cooling becomes efficient and decreases the gas temperature at densities of $n\sim 10^2$--$10^3~\mathrm{cm^{-3}}$.

The bottom panels show that the metallicity is nearly uniform throughout each collapsing cloud. The case with [Z/H]$=-10$ closely resembles the classical DCBH formation scenario, which requires nearly metal-free gas exposed to an intense LW radiation field. The other three clouds have finite metallicities and therefore do not satisfy the conventional DCBH criteria. At [Z/H]$\leq -3.0$, however, the thermal evolution remains almost identical to that of the metal-free model, producing a similar inflow structure on parsec scales.

Even at [Z/H]$=-1.9$, where the gas temperature decreases rapidly at $n\sim 10^3~\mathrm{cm^{-3}}$, the strong inflow still produces a star particle classified as an SMS in our subgrid model. The intense local radiation field maintains a large Jeans mass and delays cloud collapse until a gas mass of $10^5$--$10^6~M_\odot$ has accumulated, thereby enabling the formation of an SMS-classified object even in
metal-enriched gas \citep{Chon+2020, Chon+2025}. We note, however, that the physical resolution of the simulation is approximately $10~\mathrm{pc}$. Unresolved fragmentation below this scale may suppress the formation of a single SMS and instead lead to the formation of a dense star cluster. We return to this issue in Section~\ref{sec:discussion1}.

\subsubsection{Connection to host halo property}
Fig.~\ref{fig:BH_occupation_fraction} shows the halo occupation fraction of BHs with masses larger than $10^4~M_\odot$ and $10^5~M_\odot$, shown by the red and black lines, respectively. We here assume SMSs directly collapse into BHs. To calculate the occupation fraction, we compile halos from all snapshots and compute the fraction of halos hosting at least one BH above the given mass threshold. Because halos from different snapshots are combined, the occupation fraction reflects both the dependence on halo mass and the redshift evolution of the seed population.

The occupation fraction increases with halo mass and approaches unity above $M_{\rm halo}\sim 3\times 10^8~M_\odot$ for $M_{\rm BH}>10^4~M_\odot$ and $M_{\rm halo}\sim 10^9~M_\odot$ for $M_{\rm BH}>10^5~M_\odot$. These results support the BH seeding prescription based on the halo threshold mass, which is often adopted in the large-scale galaxy formation simulation.

Although this trend is shown as a function of halo mass, the occupation fraction is not necessarily controlled by halo mass alone. More massive halos in our simulation are typically located in clustered regions, where nearby star-forming halos enhance the LW radiation field. The globally averaged LW intensity increases as the most massive halo grows (Fig.~\ref{fig:evolution}). The high occupation fraction at large halo masses therefore reflects not only the growth of the host halo itself, but also the biased environment, where the halos and star forming galaxies are clustered and strong LW irradiation is more readily produced.

\begin{figure}[t]
\centering
\includegraphics[width=0.45\textwidth]{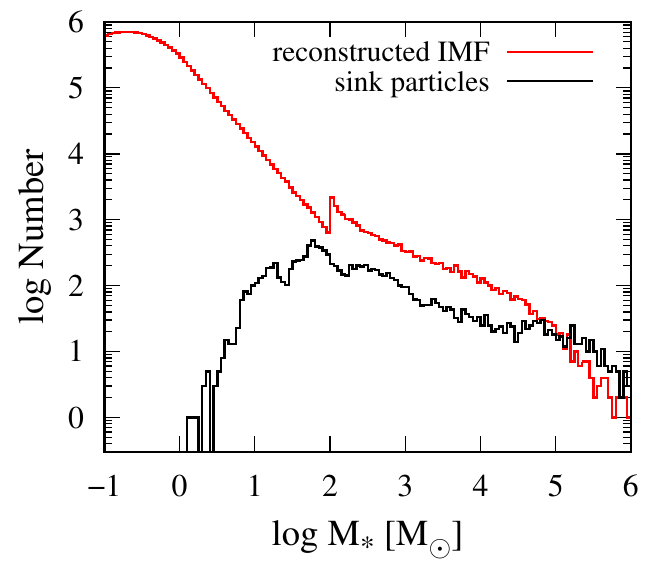}
\caption{
The black histogram shows the mass function of star particles formed in the simulation. The red histogram shows the reconstructed IMF inferred from the simulation output, assuming that 10 per cent of the stellar mass follows a Chabrier IMF and that the remaining mass is distributed according to a log-flat IMF above $100~M_\odot$.
}\label{fig:IMF}
\end{figure}

\subsubsection{IMF from Pop III to SMS regime}
\label{sec:IMF}
The black histogram in Fig.~\ref{fig:IMF} shows the mass distribution of star particles formed in the simulation. The distribution peaks at $\sim 100~M_\odot$, similar to the characteristic mass found in previous studies of Pop III star formation \citep{Stacy+2013, Hirano+2014, Hirano+2015, Sugimura+2023, Klessen+2023}. Although our sample also includes stars formed from metal-enriched gas, the thermal evolution at low metallicities remains similar to that in primordial gas, so that their characteristic stellar masses are expected to retain Pop III-like properties \citep{Chon+2021}. Above this peak, however, the distribution extends continuously toward the SMS regime, reaching stellar masses of $10^5$--$10^6~M_\odot$. Such a high-mass tail is not seen in typical Pop III star formation simulations, where the mass distribution usually shows a sharp decline above $\sim 10^3~M_\odot$ \citep{Hirano+2015}. In our simulation, this extension toward the SMS regime is produced by the strong LW radiation field, as shown in Fig.~\ref{fig:J21_Mstar}. The LW radiation suppresses H$_2$ cooling in dense gas and allows some collapsing clouds to avoid ordinary fragmentation, leading to the formation of SMSs.

The red distribution in Fig.~\ref{fig:IMF} is not a direct prediction of the cosmological simulation but a reconstructed IMF, where we account for unresolved fragmentation below our numerical resolution (Section~\ref{sec:reconstruction}). We assume that 10 per cent of the stellar mass is distributed according to a Chabrier IMF \citep{Chabrier2003} from $0.1$ to $100~M_\odot$, while the remaining mass follows a log-flat distribution from $100~M_\odot$ up to the mass of the corresponding star particle \citep{Chon+2022, Chon+2025}. This produces a discontinuity at $M_*=100~M_\odot$, because the low-mass Chabrier component and the high-mass log-flat component are treated separately. Despite this artificial feature, the reconstructed IMF retains a broad high-mass tail extending up to $\sim 10^6~M_\odot$. We use this reconstructed IMF in the following analysis of the cluster spectra.

\begin{figure} [t]
\centering
\includegraphics[width=0.5\textwidth]{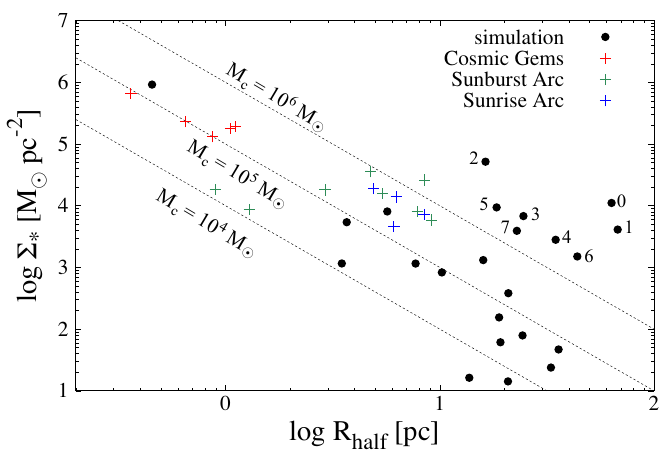}
\caption{
Stellar surface density as a function of half-mass radius, $R_{\rm half}$, for the simulated clusters, shown by black circles. The dashed lines indicate constant cluster masses of $10^4$, $10^5$, and $10^6~M_\odot$. The eight most massive clusters are labeled with cluster IDs from 0 to 7. We also show the surface densities and half-mass radii of observed strongly lensed clusters, such as, Sunburst Arc \citep{Vanzella+2022}, Sunrise Arc \citep{Vanzella+2023}, and Cosmic Gems \citep{Adamo+2024}.
}\label{fig:mass_size_relation}
\end{figure}

\begin{figure} [t]
\centering
\includegraphics[width=0.5\textwidth]{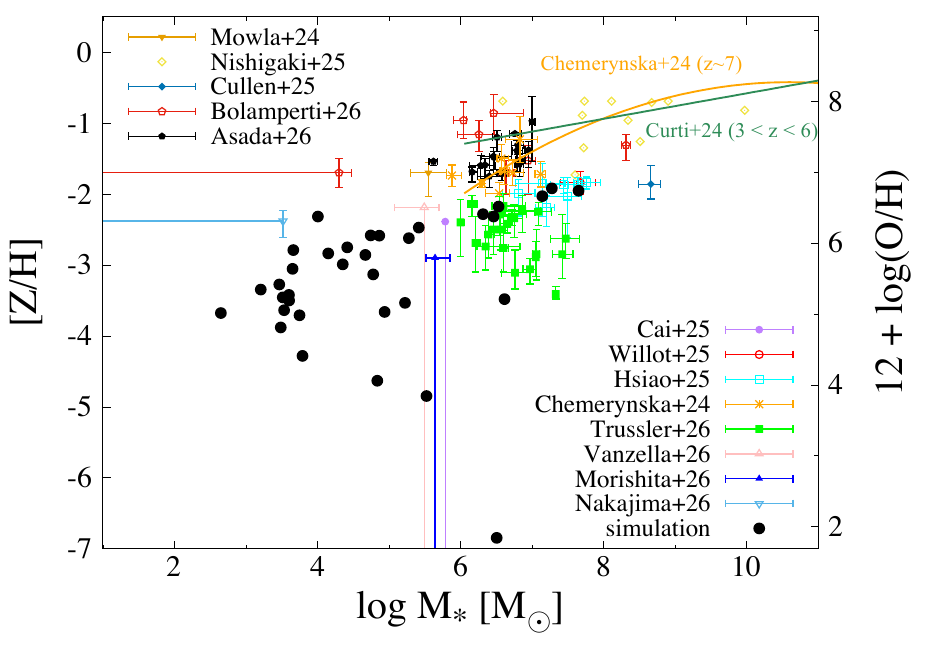}
\caption{
Mass--metallicity relation of the simulated clusters, shown by black circles.  Note that cluster IDs increase with decreasing stellar mass, with Cluster 0 being the most massive. We overplot the mass--metallicity relations of high-$z$ clusters \citep{Chemerynska+2024, Mowla+2024, Cullen+2025, Hsiao+2025, Willott+2025, Nishigaki+2025, Trussler+2026, Bolamperti+2026, Asada+2026} and extremely metal-poor clusters \citep{Cai+2025, Morishita+2025, Vanzella+2026, Nakajima+2026}. The solid lines show the relations at the high stellar-mass end \citep{Chemerynska+2024, Curti+2024}.
}\label{fig:mass_Z_relation}
\end{figure}

\subsection{Formation of compact clusters with SMS formation}
We study the morphology of the star clusters and their spectra in this section. Fig.~\ref{fig:overall} also shows the spatial distribution of star particles in four of the most massive clusters. The color indicates the stellar surface density, calculated from the mass enclosed within the radius containing the nearest 32 star particles. Clusters 0 and 1 show highly clumpy structures, with local surface density maxima of $10^3$--$10^4~M_\odot~{\rm pc}^{-2}$, whereas clusters 2 and 3 appear to be dominated by a single component.

The green asterisks mark the positions of living SMSs with masses larger than $10^4~M_\odot$. The SMS positions broadly follow the underlying stellar density structures, and they are often located near the centers of local density maxima. This indicates that SMSs contribute significantly to the local enhancement of the stellar surface density. Some SMSs, however, reside near the outskirts of their host clusters. For example, cluster 1 contains two SMSs at the edge of the cluster, separated from the nearest local density maxima by $\sim 100$ pc.

\subsubsection{Cluster statistics} \label{sec:cluster_statistics}
Fig.~\ref{fig:mass_size_relation} shows the half-mass radius, $R_{\rm half}$, and the mean stellar surface density within $R_{\rm half}$ for all 34 clusters at the final snapshot of $z=16.7$. We define $R_{\rm half}$ as the radius measured from the cluster center of mass that encloses half of the total stellar mass. Except for one outlier, the clusters have $R_{\rm half}\simeq 3$--$100$ pc. The colored symbols show observed high-$z$ lensed star clusters \citep{Vanzella+2022, Vanzella+2023, Adamo+2024}.
The most compact clusters in our simulation have sizes and surface densities comparable to those of the observed clusters in the Sunburst Arc and Sunrise Arc, while many simulated clusters are more extended. Clusters 0 and 1 are among the most extended systems in our sample, with half-mass radii of $\sim 70\,\mathrm{pc}$, comparable to those of compact galaxies at high redshift \citep{Harikane+2025}. We note, however, that these massive clusters contain significant substructure, including compact density maxima with sizes of $\lesssim 10$ pc (see Fig.~\ref{fig:overall}).

Cluster 10 is an exceptionally compact system with $R_{\rm half}\simeq 0.3$ pc. This cluster consists of two SMSs and 19 lower-mass stars. Because the two SMSs dominate the total stellar mass and are separated by only $\sim 0.3$ pc, the estimated half-mass radius is essentially set by their separation. If such SMSs are accompanied by a compact population of lower-mass stars, this system may represent a possible counterpart of the observed lensed compact clusters. We return to this point in Section~\ref{sec:discussion3}.

Fig.~\ref{fig:mass_Z_relation} shows the mass--metallicity relation of the simulated clusters at the final snapshot, shown by black circles. The cluster metallicities are typically below one per cent of the solar value and range from [Z/H]$=-5$ to $-2$, except for one cluster. The colored symbols with error bars show the mass--metallicity relation of observed clusters or cluster candidates \citep{Chemerynska+2024, Curti+2024, Hsiao+2025, Willott+2025, Trussler+2026}.
The simulated clusters with masses above $10^6~M_\odot$ have masses and metallicities comparable to those of the observed systems. Below this mass, the metallicity gradually decreases toward [Z/H]$\sim -4$, with a large scatter. These low-mass clusters are also consistent with the observed extremely metal-poor cluster \citep{Cai+2025, Morishita+2025, Vanzella+2026, Nakajima+2026}.

The most metal-poor cluster in our simulation is Cluster 5, with a metallicity of [Z/H]$\sim -7$. This system episodically forms SMSs between $z=19$ and $17$. Metal enrichment proceeds slowly because most massive stars directly collapse into BHs without ejecting metals into the surrounding gas.  This system may represent a high-redshift analogue of nearly metal-free cluster candidates such as LAP-1B \citep{Nakajima+2026} and AMORE \citep{Morishita+2025}.

\begin{figure}
\centering
\includegraphics[width=0.45\textwidth]{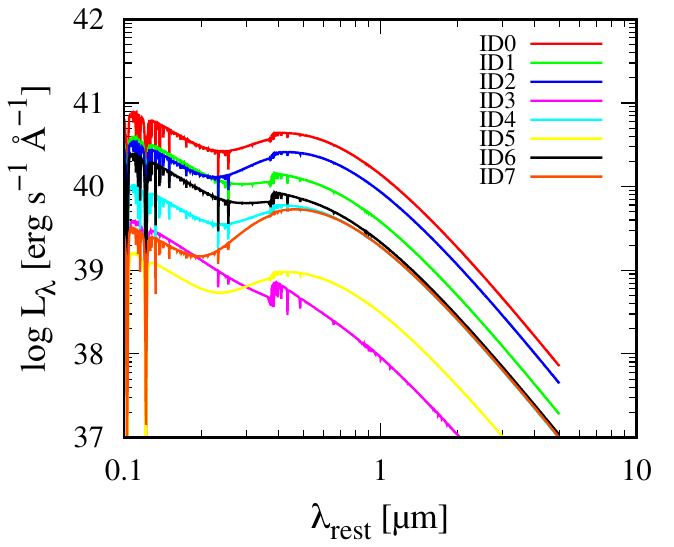}
\caption{
Spectra of the eight most massive star clusters in the simulation as a function of rest-frame wavelength, $\lambda_{\rm rest}$. 
We decompose the spectra into contributions from normal stars and SMSs. The spectra of normal stars are computed using a stellar library constructed from the MIST models and the PHOENIX stellar-atmosphere library.  The spectra of SMSs are modeled as blackbody emission with $T_{\rm eff}=6000\,{\rm K}$ and a bolometric luminosity equal to the Eddington luminosity.
}\label{fig:cluster_spectra}
\end{figure}

\begin{figure}
\centering
\includegraphics[width=0.45\textwidth]{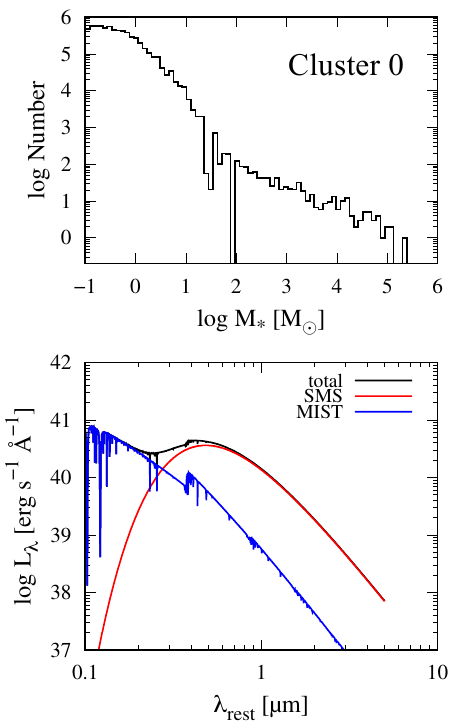}
\caption{
Example of a cluster with a prominent red optical component. 
The top panel shows the mass distribution of the living stars in Cluster 0. 
The bottom panel shows the cluster spectrum as a function of rest-frame wavelength, $\lambda_{\rm rest}$. 
The black line shows the total spectrum, while the red and blue lines show the contributions from SMSs and from the normal stellar component constructed with the MIST-based library, respectively.
}\label{fig:cluster_ID000}
\end{figure}

\begin{figure}
\centering
\includegraphics[width=0.45\textwidth]{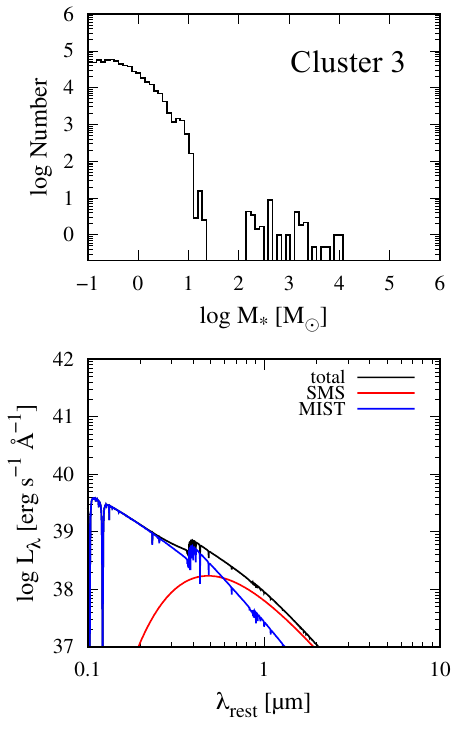}
\caption{
The same as Fig.~\ref{fig:cluster_ID000} but for Cluster 3, which has less significant optical component.
}\label{fig:cluster_ID003}
\end{figure}

\begin{figure}
\centering
\includegraphics[width=0.45\textwidth]{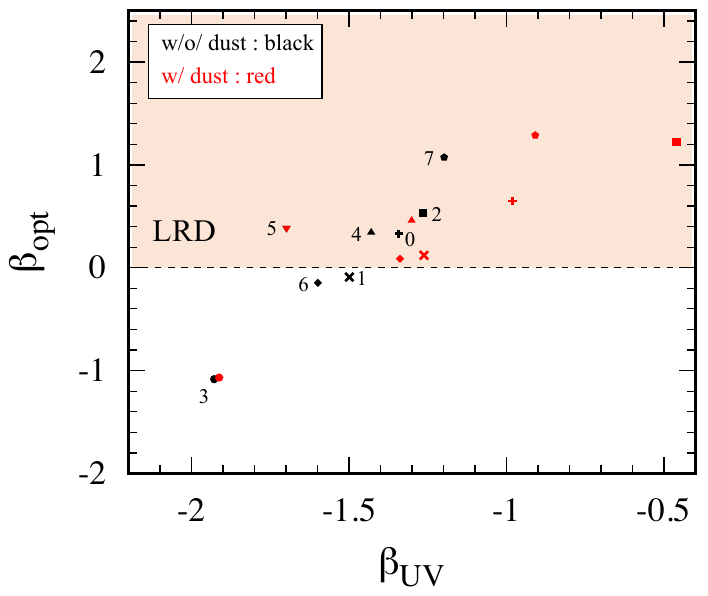}
\caption{
Slopes of the cluster spectra in the UV ($0.15$--$0.3\,\mu$m) and optical ($0.3$--$0.6\,\mu$m) bands. The black and red symbols indicate the slopes without and with dust attenuation, respectively. Different symbols distinguish individual clusters, with the same symbol used for each cluster with and without dust attenuation.
The symbols are labeled with the corresponding cluster IDs.
The hatched region with $\beta_{\mathrm{opt}}>0$ and
$\beta_{\mathrm{UV}}<0$ indicates the region occupied by V-shaped,
LRD-like spectra \citep{Setton+2025,Lin+2026}.
}\label{fig:beta_slopes}
\end{figure}

\subsubsection{Cluster spectra} \label{sec:spectra}
Fig.~\ref{fig:cluster_spectra} shows the spectra of the eight most massive clusters from Cluster ID 0 to 7, which have the stellar mass larger than $10^6~M_\odot$. Some of the clusters, in particular ID 0, ID 2, and ID 4, show a prominent red optical component together with a blue UV continuum, producing a V-shaped continuum similar to that observed in LRDs \citep[e.g.,][]{Kocevski+2023}. Other clusters, such as ID 3 and ID 6, show only a mild excess in the red optical component.

Figs.~\ref{fig:cluster_ID000} and \ref{fig:cluster_ID003} show the mass distributions of the living stars (top panels) and the corresponding spectra (bottom panels) for Cluster ID 0 and ID 3, which provide examples of an LRD-like cluster and a more normal star-forming clump, respectively. Cluster 0 is undergoing SMS formation, producing an extended log-flat tail in the stellar mass distribution. The combined bolometric luminosity of its SMS population, estimated from the adopted Eddington luminosities, is $3.46\times10^{44}~\mathrm{erg~s^{-1}}$, comparable to the typical bolometric luminosities inferred for LRDs \citep{Umeda+2026}. The spectral decomposition shown in the bottom panel demonstrates that the rise in $L_{\lambda}$ above $0.3~\mu{\rm m}$ is dominated by the SMS component.

In contrast, Cluster 3 shows a much weaker optical excess. In this case, only two SMSs with masses of $\sim 10^4~M_\odot$ contribute to the red component. The sharp rise in the optical band is partly produced by the SMSs, but the strong Balmer break in the normal stellar population also contributes. This indicates that a V-shaped continuum is not necessarily produced by a single physical component: it can arise from a combination of blue UV emission from normal massive stars, a Balmer break in the stellar population, and red optical emission from bloated SMSs. Nevertheless, the clusters with the most prominent red optical components are those in which massive SMSs dominate the optical luminosity. Thus, ongoing SMS formation provides a natural channel for producing compact, luminous, LRD-like spectra in our simulations.

To compare our sample more quantitatively to the LRD observation, we fit the cluster spectra by double power-law following \citet{Setton+2025},
\begin{align}
L_\lambda = L_{\lambda,\mathrm{break}}
\begin{cases}
\left(\lambda/\lambda_{\mathrm{break}}\right)^{\beta_{\mathrm{UV}}}, 
& \lambda \leq \lambda_{\mathrm{break}}, \\
\left(\lambda/\lambda_{\mathrm{break}}\right)^{\beta_{\mathrm{opt}}}, 
& \lambda > \lambda_{\mathrm{break}},
\end{cases}
\end{align}
where the fitting is performed for the three parameters $\beta_\text{opt}$, $\beta_\text{UV}$, and $\lambda_\text{break}$ in the spectrum at $0.15~\mu\mathrm{m} \leq \lambda \leq 0.6 \mu\mathrm{m}$.
The V-shaped continuum shape typical to the LRD is expressed as $\beta_\text{opt} > 0$ and $\beta_\text{UV} < 0$. Fig.~\ref{fig:beta_slopes} shows our fitted results of $\beta_\text{UV}$ and $\beta_\text{opt}$ for our eight most massive clusters, which host SMSs. The black crosses show the results for the spectra with no dust attenuation, indicating five clusters have V-shaped spectra. Positive $\beta_\text{opt}$ is created by the SMS component, as shown in Fig.~\ref{fig:cluster_spectra}. Clusters 1, 3 and 6 are classified as non-LRDs due to smaller contribution of SMSs.

The red symbols show the slopes including dust attenuation. We calculate the spherically averaged dust column-density profile, assuming that the dust-to-gas mass ratio scales linearly with gas metallicity. We then convert the dust column density into attenuation using the Calzetti law \citep{Calzetti+2000}.  
The dust attenuation causes reddening and makes the slope shallower. Clusters 1 and 6 satisfy the LRD criterion only after dust attenuation is included.

\section{Discussion} \label{sec:discussion}
\subsection{Pathways to heavy seed formation in metal-enriched stellar systems} \label{sec:discussion1}
Our simulation resolves the formation of dense, strongly irradiated star-forming clumps on scales of $\sim 10$ physical pc. However we do not resolve fragmentation, stellar mergers, or protostellar evolution on sub-pc scales. Therefore, the mapping between a massive star particle and a final BH seed is not unique, and we do not model subsequent gas accretion onto BHs in this work. Nevertheless, the formation of massive, compact stellar systems in weakly metal-enriched gas is a robust outcome of our simulation. As we discuss below, sub-resolution fragmentation may alter the pathway to heavy seed formation rather than simply suppress it, allowing either the growth of a dominant very massive star or merger-driven growth in a dense stellar cluster. We therefore examine below how unresolved small-scale physics affects the fate of these systems and the formation of heavy seed BHs.

In our simulation, each star particle is treated as an individual massive star. In the spectral analysis in Section~\ref{sec:spectra}, we relax this assumption and reconstruct the unresolved stellar mass distribution by allowing sub-resolution fragmentation. Following the high-resolution simulations of \citet{Chon+2025}, we assume that 10 per cent of the stellar mass is distributed according to a Chabrier IMF, while the remaining 90 per cent follows a log-flat IMF. This prescription still allows very massive individual stars to form. Our Monte Carlo sampling shows that, for a star particle of mass $M_{\rm star}$, the median mass of the most massive sampled star is $\sim 0.45 M_{\rm star}$. Thus, even after accounting for unresolved fragmentation, our model assumes that a large fraction of each massive star particle can be dominated by a single very massive star.

Recent studies have shown that SMSs can form in metal-enriched environments. Previous calculations by \citet{Chon+2020} and \citet{Chon+2025} showed that SMSs, and hence heavy seed BHs, can form at finite metallicities of [Z/H]$\lesssim -3$. In such environments, fine-structure line cooling and dust thermal emission promote small-scale fragmentation \citep[e.g.,][]{Li+2003,Tsuribe+2006,Omukai+2008,Dopcke+2013, Chon+2021}. However, massive converging flows can still feed the central objects efficiently, allowing the most massive stars to continue growing. The resulting systems develop a broad, approximately log-flat high-mass IMF, in which the most massive stars dominate the total stellar mass.

The case of [Z/H]$=-2$ appears to be transitional. In \citet{Chon+2025}, the mass of the central SMS is reduced by about an order of magnitude compared to lower-metallicity cases. At this metallicity, fine-structure line cooling induces fragmentation on scales of $\sim 1$--$10$ pc, and the system evolves into a compact star cluster with a stellar surface density of $\Sigma_* \sim 10^3$--$10^4~M_\odot\,{\rm pc^{-2}}$. Nevertheless, very massive stars with masses of $\sim 2000~M_\odot$ still form at the cluster center.

The final mass of the most massive star at [Z/H]$=-2$ should depend on the adopted stellar evolution model. \citet{Nandal+2026} showed that stellar envelopes can inflate more easily at higher metallicity, reducing the effective temperature and the UV emissivity of very massive accreting stars. This weakens radiative feedback and will help SMSs and heavy seed BHs form even in environments with [Z/H]$\sim -2$. Therefore, our simulation suggests that clustered, metal-enriched environments can provide viable sites for heavy seed formation, while the resulting seed masses and formation efficiency remain sensitive to unresolved fragmentation, stellar mergers, and the evolution of rapidly accreting very massive stars.

The total stellar mass, or equivalently the star formation efficiency inside a sink particle, is expected to be less sensitive to sub-resolution fragmentation, especially at the massive end. This is because the gas associated with massive sink particles is already highly bound on the sink scale. The characteristic stellar surface density and escape velocity are
\begin{align}
\Sigma_* &= \frac{M_*}{\pi R_\text{sink}^2} 
= 318~M_\odot~{\rm pc^{-2}}
\left( \frac{M_*}{10^5~M_\odot} \right)
\left( \frac{R_\text{sink}}{10~{\rm pc}} \right)^{-2}, \\
v_\text{esc} &= \sqrt{\frac{2GM_*}{R_\text{sink}}} 
= 9.3~{\rm km~s^{-1}}
\left( \frac{M_*}{10^5~M_\odot} \right)^{1/2}
\left( \frac{R_\text{sink}}{10~{\rm pc}} \right)^{-1/2}.
\end{align}
Thus, a star particle with $M_* \gtrsim 3\times 10^5~M_\odot$ corresponds to a system with $\Sigma_* \sim 10^3~M_\odot~{\rm pc^{-2}}$ and $v_\text{esc} \gtrsim 10~{\rm km~s^{-1}}$ on the sink scale. Such a high surface density and escape velocity imply that the cloud can remain gravitationally bound even after photoionization, allowing a high star formation efficiency \citep{He+2019,Fukushima+2021}. Therefore, although sub-resolution fragmentation can modify the IMF and reduce the UV emissivity of the stellar population, it is unlikely to strongly change the total stellar mass formed inside the most massive sink particles \citep{Fukushima+2023,Menon+2024}.

Sub-resolution fragmentation may also open an alternative pathway to heavy seed formation. If a massive sink particle fragments into a compact stellar cluster rather than forming a single dominant SMS, the high stellar density can trigger runaway stellar collisions and mergers, allowing the most massive stars to grow dynamically \citep{PZ+2004, Pacucci+2025}. This process can produce a very massive merger product or an SMS that subsequently collapses into a heavy seed BH \citep{Reinoso+2023, Fujii+2024, Rantala+2024, Chon+2025}. In this sense, fragmentation may change the dominant pathway to heavy seed formation, from direct growth of a single central object to merger-driven growth in a dense cluster, rather than simply suppressing heavy seed formation.

One caveat in treating a star particle as a compact star cluster is the treatment of SN feedback. If sub-resolution fragmentation occurs, a single star particle should contain a larger number of stars that eventually explode as core-collapse SNe or PISNe. For example, for star particles with masses of $10^5$, assuming that 10 per cent of the mass follows a Chabrier IMF and that the remaining mass follows a log-flat IMF, we expect $\sim 100$ core-collapse SNe and $\sim 0.6$ PISNe. Such additional SN feedback can reduce the star formation efficiency inside the halo or star-forming region, and can also modify the cloud morphology and metallicity evolution. However, in extremely dense environments, SN feedback may not disrupt the entire star-forming system, because most of the gas is already highly bound and the radiative losses of SN remnants can be efficient \citep{Dekel+2023}. 
Therefore, our current treatment likely underestimates the impact of SN feedback on the internal structure and enrichment history of the compact clusters, while the formation of a massive bound stellar system itself may be more robust. A more self-consistent treatment requires simulations that resolve sub-pc fragmentation and follow the clustered sequence of SNe in dense, metal-enriched gas.

\subsection{Origin of compact star clusters and SMSs} \label{sec:discussion2}
The formation of compact star clusters and SMSs in our simulation has a common physical origin: the suppression of H$_2$ cooling by intense local LW radiation. Once a strong LW field is established, H$_2$ cooling is suppressed and the gas temperature remains close to the atomic-cooling value, $\sim 8000\,{\rm K}$, up to densities of $n\lesssim 10^2$--$10^4\,{\rm cm^{-3}}$ \citep{Chon+2025}. The corresponding Jeans mass remains large, of order $10^5$--$10^6\,M_\odot$, so that a substantial amount of gas can accumulate before the onset of star formation. When this accumulated gas eventually becomes gravitationally unstable, star formation proceeds rapidly within the dense central region, producing a massive and compact stellar system as well as SMSs.

\citet{Sugimura+2024} showed that strong LW feedback delays the onset of star formation. The increased Jeans mass makes star formation more intermittent and bursty, resulting in the formation of dense and compact clusters. The suppression of early fragmentation also maintains large-scale inflows toward the densest regions, providing the high accretion rates required for the growth of very massive stars and SMSs \citep{Chon+2020,Chiaki+2023,Chon+2025}.
Our results are consistent with this picture: the LW field suppresses early fragmentation, allows massive gas clumps to assemble, and thereby promotes both compact cluster formation and the growth of very massive stars.

A similar picture has emerged in recent cosmological simulations of heavy seed formation. \citet{Prole+2026} adopted a BH seeding prescription based on the local mass accretion rate and metallicity, effectively identifying regions where the conditions are favorable for SMS formation. A later analysis showed that some Pop II clusters identified in the original calculation should host heavy seed BHs \citep{Prole+2026b}. These systems form in rapidly growing halos, where massive gas inflows provide the high accretion rates required for SMS growth. Allowing such SMS and heavy seed formation in metal-enriched regions greatly increases the predicted heavy-seed number density to $\sim 100~{\rm Mpc^{-3}}$. This is broadly consistent with our picture, in which strong inflows promote SMS formation in clustered, metal-enriched environments. We note, however, that their simulations include only a global LW background and do not account for local LW radiation from nearby star-forming regions. Including this local contribution may further enhance SMS and heavy seed formation in such environments.

\subsection{Connection to observations of compact star clusters and LRDs} \label{sec:discussion3}
The compact stellar systems produced in our simulation have properties comparable to high-redshift compact clusters and LRDs recently observed. Some simulated clusters reach stellar surface densities of $\Sigma_* \sim 10^4$--$10^5~M_\odot\,{\rm pc^{-2}}$, comparable to the compact structures observed in the Sunburst Arc \citep{Vanzella+2022} and the Sunrise Arc \citep{Vanzella+2023}. The Cosmic Gems clusters \citep{Adamo+2024} have even smaller sizes and higher surface densities, and only one of our simulated clusters reaches a comparable compactness. Since these compact systems are close to the resolution limit of our simulation, higher-resolution calculations are required to determine whether unresolved substructures can produce even denser clusters. Our simulated systems therefore provide promising candidates for follow-up studies of compact cluster formation in strongly irradiated, low-metallicity environments.

Many observed compact clusters do not show red colors comparable to those of LRDs. This can be understood as a consequence of the short duration of the bloated SMS phase. SMSs live only for $\sim 2~{\rm Myr}$ before collapsing into BHs. In addition, even before their collapse, the stellar envelopes can contract on the Kelvin--Helmholtz timescale once the accretion rate drops below $\sim 0.02~M_\odot\,{\rm yr^{-1}}$ \citep{Sakurai+2015}. The stars then evolve toward main-sequence-like structures with higher effective temperatures, making their spectra bluer. The red optical phase associated with bloated SMSs is therefore expected to be transient. After several Myr, SMSs collapse into BHs, while SNe from massive stars can disrupt the remaining gas reservoir and suppress further SMS formation and accretion. The cluster spectrum then becomes dominated by normal stellar populations and becomes bluer. This timescale effect may explain why compact clusters and LRD-like sources can be physically related, even though only a subset of compact clusters are observed in a red phase.

The clustered nature of SMS formation also provides a possible connection to the ``connecting dots'' \citep{Baggen+2026} and ``dual dots'' \citep{Tanaka+2024, Barger+2026} morphologies reported for some LRD systems. In our simulation, compact clusters form in clustered star-forming regions, where nearby stellar systems generate an intense LW radiation field and provide favorable sites for SMS formation. The typical separations between these clusters are of order a physical kpc, comparable to the separations reported in observations of multiple-component LRD systems \citep{Baggen+2026}. If more than one compact cluster hosts bloated SMSs at the same time, such a system would appear as multiple compact red components embedded in, or associated with, a larger star-forming complex.

On smaller spatial scales, local LW radiation within an individual star-forming complex can also promote SMS formation.
We find that SMSs do not necessarily form at the center of a relaxed stellar system, but can instead form in off-centered dense clumps embedded in an extended star-forming complex. For example, Cluster 1 extends over a few hundred pc and contains SMS-forming clumps near the edge of the cluster. These clumps are separated from each other by $\lesssim 100~{\rm pc}$. If SMSs appear as red optical sources, this provides a scenario in which LRD-like systems form at the outskirts of galaxies or extended star-forming complexes. This geometry is qualitatively consistent with the strongly lensed systems reported by the JWST/VENUS program \citep{Yanagisawa+2026}, where two off-centered LRDs are found near the edge of a galaxy and are separated from each other by $\sim 70~{\rm pc}$.

One caveat of our LRD model is that it does not naturally explain the large velocity widths of the observed H$\alpha$ emission lines. The broad widths may arise from microturbulence near the stellar surface \citep{Nandal&Loeb2026} or from rotational motion in the immediate vicinity of the SMSs \citep{Zwick+2026}. Another possibility is that some SMSs collapse into BHs during hydrogen burning \citep{Fuller+1986,Shibata+2002,Umeda+2016,Haemmerle+2018}, leading to a quasi-star-like configuration \citep{Begelman+2008}. If some of the SMSs evolve into quasi-stars, the resulting systems could be associated with broad H$\alpha$ emission powered by gas close to the embedded BHs. SMSs can also collapse into BHs after their lifetimes if their masses exceed $260\,M_\odot$ \citep{Heger+2002}. In this case, subsequent accretion onto the newly formed BHs could also produce broad H$\alpha$ emission.

\subsection{Connection to large-volume cosmological simulations and the fate of heavy seed BHs} \label{sec:discussion4}
Our results provide a physical interpretation for the halo-based BH seeding prescriptions commonly adopted in large-scale cosmological simulations. Fig.~\ref{fig:BH_occupation_fraction} shows that the occupation fraction of heavy seed BHs increases rapidly with halo mass and approaches unity above $M_{\rm halo}\sim 10^9~M_\odot$ for seeds with $M_{\rm BH}>10^5\,M_\odot$. In simulations such as Illustris, IllustrisTNG, and EAGLE, BH seed particles are inserted once halos exceed a threshold mass \citep[e.g.,][]{Vogelsberger+2014,Schaye+2015,Weinberger+2017,Pillepich+2018, Bhowmick+2024}. Although such prescriptions are phenomenological, our simulation shows why a halo-mass threshold can work as a statistical proxy. Massive halos preferentially reside in biased, clustered regions where strong LW radiation, large gas inflows, and previous metal enrichment coexist. These conditions make such regions favorable sites for efficient massive-seed formation. Thus, the threshold mass used in large-volume simulations should be interpreted not as a direct physical condition for seed formation, but as a coarse-grained representation of unresolved clustered, strongly irradiated SMS-forming regions.

The subsequent fate of these seeds, however, is not captured by a seeding prescription alone. In our simulation, some heavy seeds form close to the centers of dense stellar systems, but many others form in off-centered clumps. Seeds born near the center of a massive stellar system may migrate inward through dynamical friction and eventually merge with other BHs. In contrast, off-centered seeds are not guaranteed to reach the galactic center because of their long dynamical friction timescales \citep{Latif+2018,Chon+2021a}. Previous cosmological simulations have shown that seed BHs can remain displaced from galactic centers for long periods, with their abundance depending on the host potential, merger history, seed mass, and the treatment of BH dynamics and dynamical friction \citep[e.g.,][]{Bellovary+2010,Tremmel+2018,Ricarte+2021}. Our scenario provides a formation pathway for such a population: heavy seeds are produced not only at the center of a single galaxy, but also in clustered and off-centered star-forming clumps, which will constitute a population of wandering BHs. 

If seed BHs form close to the galaxy center, they will migrate and merge with other BHs, which provides a particularly important observational test.  Previous studies have shown that the LISA detection rate and the mass and redshift distributions of massive BH mergers are sensitive to the seed mass and seed occupation fraction \citep{Sesana+2007,Klein+2016,Barausse+2020}. \citet{McCaffrey+2025} showed that heavy seed formation should be associated with strong gravitational-wave signals detectable with LISA. Comparing future LISA merger rates and mass--redshift distributions with theoretical predictions will therefore provide a way to test whether heavy seeds form efficiently in clustered, metal-enriched environments. Connecting the pc-scale formation channel studied here to large-volume simulations is essential for predicting both the central SMBH population and the wandering and merging massive BH populations in the early Universe.

\section{Summary and Conclusions} \label{sec:summary}

We have performed a cosmological radiation-hydrodynamic zoom-in simulation to study the formation of heavy seed BHs in clustered, metal-enriched star-forming environments. The simulation follows primordial and metal-enriched cooling, radiative feedback, star formation, stellar evolution, and the formation of SMSs. We also construct intrinsic spectra of the simulated compact clusters using MIST/PHOENIX-based stellar spectra for normal stars and cool blackbody spectra for bloated SMSs.

We find that intense LW radiation is generated locally in clustered star-forming regions. The LW field is highly inhomogeneous: while the volume-weighted mean intensity remains modest, dense gas around star-forming systems is exposed to much stronger radiation fields. This suppresses H$_2$ cooling, allows dense gas to collapse nearly at the atomic-cooling temperature of $\sim 10,000$ K, and promotes the formation of SMSs. Most SMSs and heavy seed BHs form in weakly metal-enriched gas with [Z/H]$\lesssim -2$, because the strongest LW radiation fields are spatially correlated with previous star formation and metal enrichment. Thus, heavy seed formation in our simulation is not restricted to environments of primordial gas, but occurs in clustered regions with metallicities around [Z/H]$\sim -2$.

The same strongly irradiated environments also produce compact star clusters. The simulated clusters have masses, sizes, surface densities, and metallicities comparable to observed high-redshift lensed compact clusters. Some clusters host bloated SMSs whose red optical emission dominates the intrinsic spectrum, producing a blue UV continuum and a prominent red optical component similar to the V-shaped continua observed in LRDs. Balmer breaks in normal stellar populations can also contribute, but the most prominent red optical components are powered by emerging SMSs. The red phase is expected to be short-lived because SMSs collapse into BHs or contract to hotter main-sequence-like structures once accretion weakens.

Our results suggest that heavy seed formation, compact cluster formation, and LRD-like continuum emission can be linked outcomes of the same physical environment: dense, weakly metal-enriched, strongly irradiated star-forming regions. A more complete comparison with LRD observations will require simulations that resolve sub-pc fragmentation and stellar mergers, follow BH accretion after SMS collapse, and include nebular emission, dust attenuation, and line radiative transfer.

\begin{acknowledgments}
We thank Zoltan Haiman for constructive discussions. This work was supported by JSPS KAKENHI Grant Number JP21H01123 and JP26K00743 (S.H.). K.C. acknowledges the support from the National Science and Technology Council, Taiwan, under grant No. NSTC 115-2112-M-001-001- and the Academia Sinica, Taiwan, under a career development award under grant No. AS-CDA-111-M04.
We conduct numerical simulation on XC50 at the Center for Computational Astrophysics (CfCA) of the National Astronomical Observatory of Japan.
\end{acknowledgments}

%




\bibliography{bibliography}{}

@ARTICLE{Draine+1996,
       author = {{Draine}, B.~T. and {Bertoldi}, Frank},
        title = "{Structure of Stationary Photodissociation Fronts}",
      journal = {\apj},
         year = 1996,
        month = sep,
       volume = {468},
        pages = {269},
          doi = {10.1086/177689},
archivePrefix = {arXiv},
       eprint = {astro-ph/9603032},
 primaryClass = {astro-ph},
       adsurl = {https://ui.adsabs.harvard.edu/abs/1996ApJ...468..269D}
}

@ARTICLE{Inayoshi_dense_gas+2025,
       author = {{Inayoshi}, Kohei and {Maiolino}, Roberto},
        title = "{Extremely Dense Gas around Little Red Dots and High-redshift Active Galactic Nuclei: A Nonstellar Origin of the Balmer Break and Absorption Features}",
      journal = {\apjl},
         year = 2025,
        month = feb,
       volume = {980},
       number = {2},
          eid = {L27},
        pages = {L27},
          doi = {10.3847/2041-8213/adaebd},
archivePrefix = {arXiv},
       eprint = {2409.07805},
 primaryClass = {astro-ph.GA},
       adsurl = {https://ui.adsabs.harvard.edu/abs/2025ApJ...980L..27I}
}

@ARTICLE{Dekel+2023,
       author = {{Dekel}, Avishai and {Sarkar}, Kartick C. and {Birnboim}, Yuval and {Mandelker}, Nir and {Li}, Zhaozhou},
        title = "{Efficient formation of massive galaxies at cosmic dawn by feedback-free starbursts}",
      journal = {\mnras},
         year = 2023,
        month = aug,
       volume = {523},
       number = {3},
        pages = {3201-3218},
          doi = {10.1093/mnras/stad1557},
archivePrefix = {arXiv},
       eprint = {2303.04827},
 primaryClass = {astro-ph.GA},
       adsurl = {https://ui.adsabs.harvard.edu/abs/2023MNRAS.523.3201D}
}

@ARTICLE{Li+2003,
       author = {{Li}, Yuexing and {Klessen}, Ralf S. and {Mac Low}, Mordecai-Mark},
        title = "{The Formation of Stellar Clusters in Turbulent Molecular Clouds: Effects of the Equation of State}",
      journal = {\apj},
         year = 2003,
        month = aug,
       volume = {592},
       number = {2},
        pages = {975-985},
          doi = {10.1086/375780},
archivePrefix = {arXiv},
       eprint = {astro-ph/0302606},
 primaryClass = {astro-ph},
       adsurl = {https://ui.adsabs.harvard.edu/abs/2003ApJ...592..975L}
}

@ARTICLE{Dopcke+2013,
       author = {{Dopcke}, Gustavo and {Glover}, Simon C.~O. and {Clark}, Paul C. and {Klessen}, Ralf S.},
        title = "{On the Initial Mass Function of Low-metallicity Stars: The Importance of Dust Cooling}",
      journal = {\apj},
         year = 2013,
        month = apr,
       volume = {766},
       number = {2},
          eid = {103},
        pages = {103},
          doi = {10.1088/0004-637X/766/2/103},
archivePrefix = {arXiv},
       eprint = {1203.6842},
 primaryClass = {astro-ph.SR},
       adsurl = {https://ui.adsabs.harvard.edu/abs/2013ApJ...766..103D}
}

@ARTICLE{Tsuribe+2006,
       author = {{Tsuribe}, Toru and {Omukai}, Kazuyuki},
        title = "{Dust-cooling-induced Fragmentation of Low-Metallicity Clouds}",
      journal = {\apjl},
         year = 2006,
        month = may,
       volume = {642},
       number = {1},
        pages = {L61-L64},
          doi = {10.1086/504290},
archivePrefix = {arXiv},
       eprint = {astro-ph/0603470},
 primaryClass = {astro-ph},
       adsurl = {https://ui.adsabs.harvard.edu/abs/2006ApJ...642L..61T}
}

@ARTICLE{Omukai+2008,
       author = {{Omukai}, K. and {Schneider}, R. and {Haiman}, Z.},
        title = "{Can Supermassive Black Holes Form in Metal-enriched High-Redshift Protogalaxies?}",
      journal = {\apj},
         year = 2008,
        month = oct,
       volume = {686},
       number = {2},
        pages = {801-814},
          doi = {10.1086/591636},
archivePrefix = {arXiv},
       eprint = {0804.3141},
 primaryClass = {astro-ph},
       adsurl = {https://ui.adsabs.harvard.edu/abs/2008ApJ...686..801O}
}

@ARTICLE{Kormendy+2013,
       author = {{Kormendy}, John and {Ho}, Luis C.},
        title = "{Coevolution (Or Not) of Supermassive Black Holes and Host Galaxies}",
      journal = {\araa},
         year = 2013,
        month = aug,
       volume = {51},
       number = {1},
        pages = {511-653},
          doi = {10.1146/annurev-astro-082708-101811},
archivePrefix = {arXiv},
       eprint = {1304.7762},
 primaryClass = {astro-ph.CO},
       adsurl = {https://ui.adsabs.harvard.edu/abs/2013ARA&A..51..511K}
}

@ARTICLE{PZ+2004,
       author = {{Portegies Zwart}, Simon F. and {Baumgardt}, Holger and {Hut}, Piet and {Makino}, Junichiro and {McMillan}, Stephen L.~W.},
        title = "{Formation of massive black holes through runaway collisions in dense young star clusters}",
      journal = {\nat},
         year = 2004,
        month = apr,
       volume = {428},
       number = {6984},
        pages = {724-726},
          doi = {10.1038/nature02448},
archivePrefix = {arXiv},
       eprint = {astro-ph/0402622},
 primaryClass = {astro-ph},
       adsurl = {https://ui.adsabs.harvard.edu/abs/2004Natur.428..724P}
}

@ARTICLE{Fujii+2024,
       author = {{Fujii}, Michiko S. and {Wang}, Long and {Tanikawa}, Ataru and {Hirai}, Yutaka and {Saitoh}, Takayuki R.},
        title = "{Simulations predict intermediate-mass black hole formation in globular clusters}",
      journal = {Science},
         year = 2024,
        month = jun,
       volume = {384},
       number = {6703},
        pages = {1488-1492},
          doi = {10.1126/science.adi4211},
archivePrefix = {arXiv},
       eprint = {2406.06772},
 primaryClass = {astro-ph.GA},
       adsurl = {https://ui.adsabs.harvard.edu/abs/2024Sci...384.1488F}
}

@ARTICLE{Rantala+2024,
       author = {{Rantala}, Antti and {Naab}, Thorsten and {Lah{\'e}n}, Natalia},
        title = "{FROST-CLUSTERS - I. Hierarchical star cluster assembly boosts intermediate-mass black hole formation}",
      journal = {\mnras},
         year = 2024,
        month = jul,
       volume = {531},
       number = {3},
        pages = {3770-3799},
          doi = {10.1093/mnras/stae1413},
archivePrefix = {arXiv},
       eprint = {2403.10602},
 primaryClass = {astro-ph.GA},
       adsurl = {https://ui.adsabs.harvard.edu/abs/2024MNRAS.531.3770R}
}

@ARTICLE{Reinoso+2023,
       author = {{Reinoso}, Basti{\'a}n and {Klessen}, Ralf S. and {Schleicher}, Dominik and {Glover}, Simon C.~O. and {Solar}, P.},
        title = "{Formation of supermassive stars in the first star clusters}",
      journal = {\mnras},
         year = 2023,
        month = may,
       volume = {521},
       number = {3},
        pages = {3553-3569},
          doi = {10.1093/mnras/stad790},
archivePrefix = {arXiv},
       eprint = {2303.07827},
 primaryClass = {astro-ph.GA},
       adsurl = {https://ui.adsabs.harvard.edu/abs/2023MNRAS.521.3553R}
}

@ARTICLE{Zhang+2025,
       author = {{Zhang}, Zijian and {Li}, Mingyu and {Oguri}, Masamune and {Lin}, Xiaojing and {Inayoshi}, Kohei and {Cerny}, Catherine and {Coe}, Dan and {Diego}, Jose M. and {Fujimoto}, Seiji and {Jiang}, Linhua and {Mahler}, Guillaume and {Matthee}, Jorryt and {Naidu}, Rohan P. and {Sharon}, Keren and {Shen}, Yue and {Zitrin}, Adi and {Abdurro'uf} and {Akins}, Hollis and {Allingham}, Joseph F.~V. and {Amor{\'\i}n}, Ricardo and {Asada}, Yoshihisa and {Atek}, Hakim and {Bauer}, Franz E. and {Brada{\v{c}}}, Maru{\v{s}}a and {Bradley}, Larry D. and {Cai}, Zheng and {Cantalupo}, Sebastiano and {Conselice}, Christopher and {Dai}, Liang and {Dayal}, Pratika and {Egami}, Eiichi and {Eisenstein}, Daniel J. and {Faisst}, Andreas L. and {Fan}, Xiaohui and {Fei}, Qinyue and {Frye}, Brenda L. and {Fudamoto}, Yoshinobu and {Furtak}, Lukas J. and {Golubchik}, Miriam and {Gonz{\'a}lez-Otero}, Mauro and {Harikane}, Yuichi and {Hsiao}, Tiger Yu-Yang and {Jim{\'e}nez-Teja}, Yolanda and {Kartaltepe}, Jeyhan S. and {Kiyota}, Tomokazu and {Koekemoer}, Anton M. and {Kohno}, Kotaro and {Kokorev}, Vasily and {Kumari}, Nimisha and {Labbe}, Ivo and {Lagos}, Claudia D.~P. and {Larison}, Conor and {Liang}, Yongming and {Lucas}, Ray A. and {Lyu}, Jianwei and {Martis}, Nicholas S. and {Magdis}, Georgios E. and {Messa}, Matteo and {Nakane}, Minami and {Noirot}, Ga{\"e}l and {Ortiz}, III, Rafael and {Ouchi}, Masami and {Pierel}, Justin D.~R. and {Postman}, Marc and {Reddy}, Naveen and {Ricotti}, Massimo and {Schaerer}, Daniel and {Schneider}, Raffaella and {Steidel}, Charles C. and {Tee}, Wei Leong and {Tripodi}, Roberta and {Trussler}, James A.~A. and {Umeda}, Hiroya and {Valentino}, Francesco and {Vanzella}, Eros and {Wang}, Feige and {Windhorst}, Rogier and {Wu}, Yunjing and {Wu}, Zihao and {Yanagisawa}, Hiroto and {Yang}, Jinyi and {Sun}, Fengwu},
        title = "{Little red dot variability over a century reveals black hole envelope via a giant Einstein cross}",
      journal = {arXiv e-prints},
         year = 2025,
        month = dec,
          eid = {arXiv:2512.05180},
        pages = {arXiv:2512.05180},
          doi = {10.48550/arXiv.2512.05180},
archivePrefix = {arXiv},
       eprint = {2512.05180},
 primaryClass = {astro-ph.GA},
       adsurl = {https://ui.adsabs.harvard.edu/abs/2025arXiv251205180Z}
}

@ARTICLE{Habouzit+2016,
       author = {{Habouzit}, M{\'e}lanie and {Volonteri}, Marta and {Latif}, Muhammad and {Dubois}, Yohan and {Peirani}, S{\'e}bastien},
        title = "{On the number density of `direct collapse' black hole seeds}",
      journal = {\mnras},
         year = 2016,
        month = nov,
       volume = {463},
       number = {1},
        pages = {529-540},
          doi = {10.1093/mnras/stw1924},
archivePrefix = {arXiv},
       eprint = {1601.00557},
 primaryClass = {astro-ph.GA},
       adsurl = {https://ui.adsabs.harvard.edu/abs/2016MNRAS.463..529H}
}

@ARTICLE{Bhowmick+2024,
       author = {{Bhowmick}, Aklant K. and {Blecha}, Laura and {Torrey}, Paul and {Weinberger}, Rainer and {Kelley}, Luke Zoltan and {Vogelsberger}, Mark and {Hernquist}, Lars and {Somerville}, Rachel S.},
        title = "{Representing low-mass black hole seeds in cosmological simulations: A new sub-grid stochastic seed model}",
      journal = {\mnras},
         year = 2024,
        month = apr,
       volume = {529},
       number = {4},
        pages = {3768-3792},
          doi = {10.1093/mnras/stae780},
archivePrefix = {arXiv},
       eprint = {2309.15341},
 primaryClass = {astro-ph.GA},
       adsurl = {https://ui.adsabs.harvard.edu/abs/2024MNRAS.529.3768B}
}

@ARTICLE{Offner+2010,
       author = {{Offner}, Stella S.~R. and {Kratter}, Kaitlin M. and {Matzner}, Christopher D. and {Krumholz}, Mark R. and {Klein}, Richard I.},
        title = "{The Formation of Low-mass Binary Star Systems Via Turbulent Fragmentation}",
      journal = {\apj},
         year = 2010,
        month = dec,
       volume = {725},
       number = {2},
        pages = {1485-1494},
          doi = {10.1088/0004-637X/725/2/1485},
archivePrefix = {arXiv},
       eprint = {1010.3702},
 primaryClass = {astro-ph.SR},
       adsurl = {https://ui.adsabs.harvard.edu/abs/2010ApJ...725.1485O}
}

@ARTICLE{Bromm+2003,
       author = {{Bromm}, Volker and {Loeb}, Abraham},
        title = "{Formation of the First Supermassive Black Holes}",
      journal = {\apj},
         year = 2003,
        month = oct,
       volume = {596},
       number = {1},
        pages = {34-46},
          doi = {10.1086/377529},
archivePrefix = {arXiv},
       eprint = {astro-ph/0212400},
 primaryClass = {astro-ph},
       adsurl = {https://ui.adsabs.harvard.edu/abs/2003ApJ...596...34B}
}

@ARTICLE{Kocevski+2023,
       author = {{Kocevski}, Dale D. and {Onoue}, Masafusa and {Inayoshi}, Kohei and {Trump}, Jonathan R. and {Arrabal Haro}, Pablo and {Grazian}, Andrea and {Dickinson}, Mark and {Finkelstein}, Steven L. and {Kartaltepe}, Jeyhan S. and {Hirschmann}, Michaela and {Aird}, James and {Holwerda}, Benne W. and {Fujimoto}, Seiji and {Juneau}, St{\'e}phanie and {Amor{\'\i}n}, Ricardo O. and {Backhaus}, Bren E. and {Bagley}, Micaela B. and {Barro}, Guillermo and {Bell}, Eric F. and {Bisigello}, Laura and {Calabr{\`o}}, Antonello and {Cleri}, Nikko J. and {Cooper}, M.~C. and {Ding}, Xuheng and {Grogin}, Norman A. and {Ho}, Luis C. and {Hutchison}, Taylor A. and {Inoue}, Akio K. and {Jiang}, Linhua and {Jones}, Brenda and {Koekemoer}, Anton M. and {Li}, Wenxiu and {Li}, Zhengrong and {McGrath}, Elizabeth J. and {Molina}, Juan and {Papovich}, Casey and {P{\'e}rez-Gonz{\'a}lez}, Pablo G. and {Pirzkal}, Nor and {Wilkins}, Stephen M. and {Yang}, Guang and {Yung}, L.~Y. Aaron},
        title = "{Hidden Little Monsters: Spectroscopic Identification of Low-mass, Broad-line AGNs at z > 5 with CEERS}",
      journal = {\apjl},
         year = 2023,
        month = sep,
       volume = {954},
       number = {1},
          eid = {L4},
        pages = {L4},
          doi = {10.3847/2041-8213/ace5a0},
archivePrefix = {arXiv},
       eprint = {2302.00012},
 primaryClass = {astro-ph.GA},
       adsurl = {https://ui.adsabs.harvard.edu/abs/2023ApJ...954L...4K}
}

@ARTICLE{Hviding+2026,
       author = {{Hviding}, Raphael E. and {de Graaff}, Anna and {Liu}, Hanpu and {Goulding}, Andy D. and {Ma}, Yilun and {Greene}, Jenny E. and {Boogaard}, Leindert A. and {Bunker}, Andrew J. and {Cleri}, Nikko J. and {Franx}, Marijn and {Hirschmann}, Michaela and {Leja}, Joel and {Matthee}, Jorryt and {Naidu}, Rohan P. and {Setton}, David J. and {{\"U}bler}, Hannah and {Venturi}, Giacomo and {Wang}, Bingjie},
        title = "{The X-Ray Dot: Exotic Dust or a Late-stage Little Red Dot?}",
      journal = {\apjl},
         year = 2026,
        month = mar,
       volume = {1000},
       number = {1},
          eid = {L18},
        pages = {L18},
          doi = {10.3847/2041-8213/ae4c88},
archivePrefix = {arXiv},
       eprint = {2601.09778},
 primaryClass = {astro-ph.GA},
       adsurl = {https://ui.adsabs.harvard.edu/abs/2026ApJ..1000L..18H}
}

@ARTICLE{Kokubo&Harikane2025,
       author = {{Kokubo}, Mitsuru and {Harikane}, Yuichi},
        title = "{Challenging the Active Galactic Nucleus Scenario for JWST/NIRSpec Little Red Dot and Non─Little Red Dot Broad H{\ensuremath{\alpha}} Emitters in Light of Nondetection of NIRCam Photometric Variability and X-Ray}",
      journal = {\apj},
         year = 2025,
        month = dec,
       volume = {995},
       number = {1},
          eid = {24},
        pages = {24},
          doi = {10.3847/1538-4357/ae119e},
archivePrefix = {arXiv},
       eprint = {2407.04777},
 primaryClass = {astro-ph.GA},
       adsurl = {https://ui.adsabs.harvard.edu/abs/2025ApJ...995...24K}
}

@ARTICLE{Chon+2016,
   author = {{Chon}, S. and {Hirano}, S. and {Hosokawa}, T. and {Yoshida}, N.},
    title = "{Cosmological Simulations of Early Black Hole Formation: Halo Mergers, Tidal Disruption, and the Conditions for Direct Collapse}",
  journal = {\apj},
archivePrefix = "arXiv",
   eprint = {1603.08923},
     year = 2016,
    month = dec,
   volume = 832,
      eid = {134},
    pages = {134},
      doi = {10.3847/0004-637X/832/2/134},
   adsurl = {http://ads.nao.ac.jp/abs/2016ApJ...832..134C}
}

@ARTICLE{Chon+2017,
   author = {{Chon}, S. and {Latif}, M.~A.},
    title = "{The impact of ionizing radiation on the formation of a supermassive star in the early Universe}",
  journal = {\mnras},
archivePrefix = "arXiv",
   eprint = {1611.08594},
     year = 2017,
    month = jun,
   volume = 467,
    pages = {4293-4303},
      doi = {10.1093/mnras/stx348},
   adsurl = {http://adsabs.harvard.edu/abs/2017MNRAS.467.4293C}
}

@ARTICLE{Chon+2018,
   author = {{Chon}, S. and {Hosokawa}, T. and {Yoshida}, N.},
    title = "{Radiation hydrodynamics simulations of the formation of direct-collapse supermassive stellar systems}",
  journal = {\mnras},
archivePrefix = "arXiv",
   eprint = {1711.05262},
     year = 2018,
    month = apr,
   volume = 475,
    pages = {4104-4121},
      doi = {10.1093/mnras/sty086},
   adsurl = {http://adsabs.harvard.edu/abs/2018MNRAS.475.4104C}
}

@ARTICLE{Sugimura+2014,
   author = {{Sugimura}, K. and {Omukai}, K. and {Inoue}, A.~K.},
    title = "{The critical radiation intensity for direct collapse black hole formation: dependence on the radiation spectral shape}",
  journal = {\mnras},
archivePrefix = "arXiv",
   eprint = {1407.4039},
     year = 2014,
    month = nov,
   volume = 445,
    pages = {544-553},
      doi = {10.1093/mnras/stu1778},
   adsurl = {http://adsabs.harvard.edu/abs/2014MNRAS.445..544S}
}

@ARTICLE{Sugimura+2023,
       author = {{Sugimura}, Kazuyuki and {Matsumoto}, Tomoaki and {Hosokawa}, Takashi and {Hirano}, Shingo and {Omukai}, Kazuyuki},
        title = "{Formation of Massive and Wide First-star Binaries in Radiation Hydrodynamic Simulations}",
      journal = {\apj},
         year = 2023,
        month = dec,
       volume = {959},
       number = {1},
          eid = {17},
        pages = {17},
          doi = {10.3847/1538-4357/ad02fc},
archivePrefix = {arXiv},
       eprint = {2307.15108},
 primaryClass = {astro-ph.CO},
       adsurl = {https://ui.adsabs.harvard.edu/abs/2023ApJ...959...17S}
}

@ARTICLE{Sugimura+2024,
       author = {{Sugimura}, Kazuyuki and {Ricotti}, Massimo and {Park}, Jongwon and {Garcia}, Fred Angelo Batan and {Yajima}, Hidenobu},
        title = "{Violent Starbursts and Quiescence Induced by Far-ultraviolet Radiation Feedback in Metal-poor Galaxies at High Redshift}",
      journal = {\apj},
         year = 2024,
        month = jul,
       volume = {970},
       number = {1},
          eid = {14},
        pages = {14},
          doi = {10.3847/1538-4357/ad499a},
archivePrefix = {arXiv},
       eprint = {2403.04824},
 primaryClass = {astro-ph.GA},
       adsurl = {https://ui.adsabs.harvard.edu/abs/2024ApJ...970...14S}
}

@ARTICLE{Susa2006,
   author = {{Susa}, H.},
    title = "{Smoothed Particle Hydrodynamics Coupled with Radiation Transfer}",
  journal = {\pasj},
   eprint = {astro-ph/0601642},
     year = 2006,
    month = apr,
   volume = 58,
    pages = {445-460},
      doi = {10.1093/pasj/58.2.445},
   adsurl = {http://ads.nao.ac.jp/abs/2006PASJ...58..445S}
}

@ARTICLE{Hasegawa+2010,
       author = {{Hasegawa}, K. and {Umemura}, M.},
        title = "{START: smoothed particle hydrodynamics with tree-based accelerated radiative transfer}",
      journal = {\mnras},
         year = 2010,
        month = oct,
       volume = {407},
       number = {4},
        pages = {2632-2644},
          doi = {10.1111/j.1365-2966.2010.17100.x},
archivePrefix = {arXiv},
       eprint = {1005.5312},
 primaryClass = {astro-ph.IM},
       adsurl = {https://ui.adsabs.harvard.edu/abs/2010MNRAS.407.2632H}
}

@ARTICLE{Hirano+2014,
   author = {{Hirano}, S. and {Hosokawa}, T. and {Yoshida}, N. and {Umeda}, H. and 
	{Omukai}, K. and {Chiaki}, G. and {Yorke}, H.~W.},
    title = "{One Hundred First Stars: Protostellar Evolution and the Final Masses}",
  journal = {\apj},
archivePrefix = "arXiv",
   eprint = {1308.4456},
 primaryClass = "astro-ph.CO",
     year = 2014,
    month = feb,
   volume = 781,
      eid = {60},
    pages = {60},
      doi = {10.1088/0004-637X/781/2/60},
   adsurl = {http://adsabs.harvard.edu/abs/2014ApJ...781...60H}
}

@ARTICLE{Umeda+2016,
       author = {{Umeda}, Hideyuki and {Hosokawa}, Takashi and {Omukai}, Kazuyuki and {Yoshida}, Naoki},
        title = "{The Final Fates of Accreting Supermassive Stars}",
      journal = {\apjl},
         year = 2016,
        month = oct,
       volume = {830},
       number = {2},
          eid = {L34},
        pages = {L34},
          doi = {10.3847/2041-8205/830/2/L34},
archivePrefix = {arXiv},
       eprint = {1609.04457},
 primaryClass = {astro-ph.SR},
       adsurl = {https://ui.adsabs.harvard.edu/abs/2016ApJ...830L..34U}
}

@ARTICLE{Hirano+2015,
   author = {{Hirano}, S. and {Hosokawa}, T. and {Yoshida}, N. and {Omukai}, K. and 
	{Yorke}, H.~W.},
    title = "{Primordial star formation under the influence of far ultraviolet radiation: 1540 cosmological haloes and the stellar mass distribution}",
  journal = {\mnras},
archivePrefix = "arXiv",
   eprint = {1501.01630},
     year = 2015,
    month = mar,
   volume = 448,
    pages = {568-587},
      doi = {10.1093/mnras/stv044},
   adsurl = {http://adsabs.harvard.edu/abs/2015MNRAS.448..568H}
}

@ARTICLE{Toyouchi+2023,
       author = {{Toyouchi}, Daisuke and {Inayoshi}, Kohei and {Li}, Wenxiu and {Haiman}, Zolt{\'a}n and {Kuiper}, Rolf},
        title = "{Radiative feedback on supermassive star formation: the massive end of the Population III initial mass function}",
      journal = {\mnras},
         year = 2023,
        month = jan,
       volume = {518},
       number = {2},
        pages = {1601-1616},
          doi = {10.1093/mnras/stac3191},
archivePrefix = {arXiv},
       eprint = {2206.14459},
 primaryClass = {astro-ph.GA},
       adsurl = {https://ui.adsabs.harvard.edu/abs/2023MNRAS.518.1601T}
}

@ARTICLE{Hosokawa+2009,
       author = {{Hosokawa}, Takashi and {Omukai}, Kazuyuki},
        title = "{Evolution of Massive Protostars with High Accretion Rates}",
      journal = {\apj},
         year = 2009,
        month = jan,
       volume = {691},
       number = {1},
        pages = {823-846},
          doi = {10.1088/0004-637X/691/1/823},
archivePrefix = {arXiv},
       eprint = {0806.4122},
 primaryClass = {astro-ph},
       adsurl = {https://ui.adsabs.harvard.edu/abs/2009ApJ...691..823H}
}

@ARTICLE{Sakurai+2015,
   author = {{Sakurai}, Y. and {Hosokawa}, T. and {Yoshida}, N. and {Yorke}, H.~W.
	},
    title = "{Formation of primordial supermassive stars by burst accretion}",
  journal = {\mnras},
archivePrefix = "arXiv",
   eprint = {1505.03954},
 primaryClass = "astro-ph.SR",
     year = 2015,
    month = sep,
   volume = 452,
    pages = {755-764},
      doi = {10.1093/mnras/stv1346},
   adsurl = {http://adsabs.harvard.edu/abs/2015MNRAS.452..755S}
}

@ARTICLE{Springel2010,
       author = {{Springel}, Volker},
        title = "{E pur si muove: Galilean-invariant cosmological hydrodynamical simulations on a moving mesh}",
      journal = {\mnras},
         year = 2010,
        month = jan,
       volume = {401},
       number = {2},
        pages = {791-851},
          doi = {10.1111/j.1365-2966.2009.15715.x},
archivePrefix = {arXiv},
       eprint = {0901.4107},
 primaryClass = {astro-ph.CO},
       adsurl = {https://ui.adsabs.harvard.edu/abs/2010MNRAS.401..791S}
}

@ARTICLE{Heger+2002,
       author = {{Heger}, A. and {Woosley}, S.~E.},
        title = "{The Nucleosynthetic Signature of Population III}",
      journal = {\apj},
         year = 2002,
        month = mar,
       volume = {567},
       number = {1},
        pages = {532-543},
          doi = {10.1086/338487},
archivePrefix = {arXiv},
       eprint = {astro-ph/0107037},
 primaryClass = {astro-ph},
       adsurl = {https://ui.adsabs.harvard.edu/abs/2002ApJ...567..532H}
}

@ARTICLE{Zwick+2026,
       author = {{Zwick}, Lorenz and {Tiede}, Christopher and {Mayer}, Lucio},
        title = "{Little Red Dots as Self-gravitating Disks Accreting on Supermassive Stars: Spectral Appearance and Formation Pathway of the Progenitors to Direct Collapse Black Holes}",
      journal = {\apj},
         year = 2026,
        month = may,
       volume = {1002},
       number = {1},
          eid = {7},
        pages = {7},
          doi = {10.3847/1538-4357/ae4d47},
archivePrefix = {arXiv},
       eprint = {2507.22014},
 primaryClass = {astro-ph.GA},
       adsurl = {https://ui.adsabs.harvard.edu/abs/2026ApJ..1002....7Z}
}

@ARTICLE{Pacucci+2025,
       author = {{Pacucci}, Fabio and {Hernquist}, Lars and {Fujii}, Michiko},
        title = "{Little Red Dots are Nurseries of Massive Black Holes}",
      journal = {\apj},
         year = 2025,
        month = nov,
       volume = {994},
       number = {1},
          eid = {40},
        pages = {40},
          doi = {10.3847/1538-4357/ae1619},
archivePrefix = {arXiv},
       eprint = {2509.02664},
 primaryClass = {astro-ph.GA},
       adsurl = {https://ui.adsabs.harvard.edu/abs/2025ApJ...994...40P}
}

@ARTICLE{Latif+2013,
       author = {{Latif}, M.~A. and {Schleicher}, D.~R.~G. and {Schmidt}, W. and {Niemeyer}, J.~C.},
        title = "{The characteristic black hole mass resulting from direct collapse in the early Universe}",
      journal = {\mnras},
         year = 2013,
        month = dec,
       volume = {436},
       number = {4},
        pages = {2989-2996},
          doi = {10.1093/mnras/stt1786},
archivePrefix = {arXiv},
       eprint = {1309.1097},
 primaryClass = {astro-ph.CO},
       adsurl = {https://ui.adsabs.harvard.edu/abs/2013MNRAS.436.2989L}
}

@ARTICLE{Latif+2014,
   author = {{Latif}, M.~A. and {Bovino}, S. and {Van Borm}, C. and {Grassi}, T. and 
	{Schleicher}, D.~R.~G. and {Spaans}, M.},
    title = "{A UV flux constraint on the formation of direct collapse black holes}",
  journal = {\mnras},
archivePrefix = "arXiv",
   eprint = {1404.5773},
     year = 2014,
    month = sep,
   volume = 443,
    pages = {1979-1987},
      doi = {10.1093/mnras/stu1230},
   adsurl = {http://adsabs.harvard.edu/abs/2014MNRAS.443.1979L}
}

@ARTICLE{Regan+2016,
   author = {{Regan}, J.~A. and {Johansson}, P.~H. and {Wise}, J.~H.},
    title = "{Forming supermassive black hole seeds under the influence of a nearby anisotropic multifrequency source}",
  journal = {\mnras},
archivePrefix = "arXiv",
   eprint = {1511.00696},
     year = 2016,
    month = jul,
   volume = 459,
    pages = {3377-3394},
      doi = {10.1093/mnras/stw899},
   adsurl = {http://ads.nao.ac.jp/abs/2016MNRAS.459.3377R}
}

@ARTICLE{Magg+2022,
       author = {{Magg}, Mattis and {Schauer}, Anna T.~P. and {Klessen}, Ralf S. and {Glover}, Simon C.~O. and {Tress}, Robin G. and {Jaura}, Ondrej},
        title = "{Metal Mixing in Minihalos: The Descendants of Pair-instability Supernovae}",
      journal = {\apj},
         year = 2022,
        month = apr,
       volume = {929},
       number = {2},
          eid = {119},
        pages = {119},
          doi = {10.3847/1538-4357/ac5aac},
archivePrefix = {arXiv},
       eprint = {2110.15372},
 primaryClass = {astro-ph.GA},
       adsurl = {https://ui.adsabs.harvard.edu/abs/2022ApJ...929..119M}
}

@ARTICLE{Sullivan+2025,
       author = {{Sullivan}, James and {Haiman}, Zolt{\'a}n and {Kulkarni}, Mihir and {Visbal}, Eli},
        title = "{Can supermassive stars form in protogalaxies due to internal Lyman─Werner feedback?}",
      journal = {\mnras},
         year = 2025,
        month = sep,
       volume = {542},
       number = {2},
        pages = {822-838},
          doi = {10.1093/mnras/staf1269},
archivePrefix = {arXiv},
       eprint = {2501.12986},
 primaryClass = {astro-ph.GA},
       adsurl = {https://ui.adsabs.harvard.edu/abs/2025MNRAS.542..822S}
}

@ARTICLE{Setton+2025,
       author = {{Setton}, David J. and {Greene}, Jenny E. and {de Graaff}, Anna and {Ma}, Yilun and {Leja}, Joel and {Matthee}, Jorryt and {Bezanson}, Rachel and {Boogaard}, Leindert A. and {Cleri}, Nikko J. and {Katz}, Harley and {Labbe}, Ivo and {Maseda}, Michael V. and {McConachie}, Ian and {Miller}, Tim B. and {Price}, Sedona H. and {Suess}, Katherine A. and {van Dokkum}, Pieter and {Wang}, Bingjie and {Weibel}, Andrea and {Whitaker}, Katherine E. and {Williams}, Christina C.},
        title = "{Little Red Dots at an Inflection Point: Ubiquitous V-shaped Turnover Consistently Occurs at the Balmer Limit}",
      journal = {\apj},
         year = 2025,
        month = dec,
       volume = {995},
       number = {1},
          eid = {118},
        pages = {118},
          doi = {10.3847/1538-4357/ae1500},
archivePrefix = {arXiv},
       eprint = {2411.03424},
 primaryClass = {astro-ph.GA},
       adsurl = {https://ui.adsabs.harvard.edu/abs/2025ApJ...995..118S}
}

@ARTICLE{Lin+2026,
       author = {{Lin}, Xiaojing and {Fan}, Xiaohui and {Cai}, Zheng and {Bian}, Fuyan and {Liu}, Hanpu and {Sun}, Fengwu and {Ma}, Yilun and {Greene}, Jenny E. and {Strauss}, Michael A. and {Green}, Richard and {Lyu}, Jianwei and {Champagne}, Jaclyn B. and {Goulding}, Andy D. and {Inayoshi}, Kohei and {Jin}, Xiangyu and {Leung}, Gene C.~K. and {Li}, Mingyu and {Liu}, Weizhe and {Liu}, Yichen and {Mao}, Junjie and {Pudoka}, Maria Anne and {Tee}, Wei Leong and {Wang}, Ben and {Wang}, Feige and {Wu}, Yunjing and {Yang}, Jinyi and {Zhang}, Haowen and {Zhu}, Yongda},
        title = "{The Discovery of Little Red Dots in the Local Universe: Signatures of Cool Gas Envelopes}",
      journal = {\apj},
         year = 2026,
        month = feb,
       volume = {997},
       number = {2},
          eid = {364},
        pages = {364},
          doi = {10.3847/1538-4357/ae2bdf},
archivePrefix = {arXiv},
       eprint = {2507.10659},
 primaryClass = {astro-ph.GA},
       adsurl = {https://ui.adsabs.harvard.edu/abs/2026ApJ...997..364L}
}

@ARTICLE{Calzetti+2000,
       author = {{Calzetti}, Daniela and {Armus}, Lee and {Bohlin}, Ralph C. and {Kinney}, Anne L. and {Koornneef}, Jan and {Storchi-Bergmann}, Thaisa},
        title = "{The Dust Content and Opacity of Actively Star-forming Galaxies}",
      journal = {\apj},
         year = 2000,
        month = apr,
       volume = {533},
       number = {2},
        pages = {682-695},
          doi = {10.1086/308692},
archivePrefix = {arXiv},
       eprint = {astro-ph/9911459},
 primaryClass = {astro-ph},
       adsurl = {https://ui.adsabs.harvard.edu/abs/2000ApJ...533..682C}
}

@ARTICLE{Dotter2016,
       author = {{Dotter}, Aaron},
        title = "{MESA Isochrones and Stellar Tracks (MIST) 0: Methods for the Construction of Stellar Isochrones}",
      journal = {\apjs},
         year = 2016,
        month = jan,
       volume = {222},
       number = {1},
          eid = {8},
        pages = {8},
          doi = {10.3847/0067-0049/222/1/8},
archivePrefix = {arXiv},
       eprint = {1601.05144},
 primaryClass = {astro-ph.SR},
       adsurl = {https://ui.adsabs.harvard.edu/abs/2016ApJS..222....8D}
}

@ARTICLE{Choi+2016,
       author = {{Choi}, Jieun and {Dotter}, Aaron and {Conroy}, Charlie and {Cantiello}, Matteo and {Paxton}, Bill and {Johnson}, Benjamin D.},
        title = "{Mesa Isochrones and Stellar Tracks (MIST). I. Solar-scaled Models}",
      journal = {\apj},
         year = 2016,
        month = jun,
       volume = {823},
       number = {2},
          eid = {102},
        pages = {102},
          doi = {10.3847/0004-637X/823/2/102},
archivePrefix = {arXiv},
       eprint = {1604.08592},
 primaryClass = {astro-ph.SR},
       adsurl = {https://ui.adsabs.harvard.edu/abs/2016ApJ...823..102C}
}

@ARTICLE{Shang+2010,
   author = {{Shang}, C. and {Bryan}, G.~L. and {Haiman}, Z.},
    title = "{Supermassive black hole formation by direct collapse: keeping protogalactic gas H$_{2}$ free in dark matter haloes with virial temperatures T$_{vir}$ $\gtrsim$ 10$^{4}$ K}",
  journal = {\mnras},
archivePrefix = "arXiv",
   eprint = {0906.4773},
 primaryClass = "astro-ph.CO",
     year = 2010,
    month = feb,
   volume = 402,
    pages = {1249-1262},
      doi = {10.1111/j.1365-2966.2009.15960.x},
   adsurl = {http://adsabs.harvard.edu/abs/2010MNRAS.402.1249S}
}

@ARTICLE{Chang+2025,
       author = {{Chang}, Seok-Jun and {Gronke}, Max and {Matthee}, Jorryt and {Mason}, Charlotte},
        title = "{Impact of resonance, Raman, and Thomson scattering on hydrogen line formation in Little Red Dots}",
      journal = {\mnras},
         year = 2026,
        month = feb,
       volume = {545},
       number = {4},
          eid = {staf2131},
        pages = {staf2131},
          doi = {10.1093/mnras/staf2131},
archivePrefix = {arXiv},
       eprint = {2508.08768},
 primaryClass = {astro-ph.GA},
       adsurl = {https://ui.adsabs.harvard.edu/abs/2026MNRAS.545f2131C}
}

@ARTICLE{Ubler+2023,
       author = {{{\"U}bler}, Hannah and {Maiolino}, Roberto and {Curtis-Lake}, Emma and {P{\'e}rez-Gonz{\'a}lez}, Pablo G. and {Curti}, Mirko and {Perna}, Michele and {Arribas}, Santiago and {Charlot}, St{\'e}phane and {Marshall}, Madeline A. and {D'Eugenio}, Francesco and {Scholtz}, Jan and {Bunker}, Andrew and {Carniani}, Stefano and {Ferruit}, Pierre and {Jakobsen}, Peter and {Rix}, Hans-Walter and {Rodr{\'\i}guez Del Pino}, Bruno and {Willott}, Chris J. and {Boeker}, Torsten and {Cresci}, Giovanni and {Jones}, Gareth C. and {Kumari}, Nimisha and {Rawle}, Tim},
        title = "{GA-NIFS: A massive black hole in a low-metallicity AGN at z {\ensuremath{\sim}} 5.55 revealed by JWST/NIRSpec IFS}",
      journal = {\aap},
         year = 2023,
        month = sep,
       volume = {677},
          eid = {A145},
        pages = {A145},
          doi = {10.1051/0004-6361/202346137},
archivePrefix = {arXiv},
       eprint = {2302.06647},
 primaryClass = {astro-ph.GA},
       adsurl = {https://ui.adsabs.harvard.edu/abs/2023A&A...677A.145U}
}

@ARTICLE{Harikane+2023,
       author = {{Harikane}, Yuichi and {Zhang}, Yechi and {Nakajima}, Kimihiko and {Ouchi}, Masami and {Isobe}, Yuki and {Ono}, Yoshiaki and {Hatano}, Shun and {Xu}, Yi and {Umeda}, Hiroya},
        title = "{A JWST/NIRSpec First Census of Broad-line AGNs at z = 4-7: Detection of 10 Faint AGNs with M $_{BH}$ {}10$^{6}$-{}10$^{8}$ M $_{{\ensuremath{\odot}}}$ and Their Host Galaxy Properties}",
      journal = {\apj},
         year = 2023,
        month = dec,
       volume = {959},
       number = {1},
          eid = {39},
        pages = {39},
          doi = {10.3847/1538-4357/ad029e},
archivePrefix = {arXiv},
       eprint = {2303.11946},
 primaryClass = {astro-ph.GA},
       adsurl = {https://ui.adsabs.harvard.edu/abs/2023ApJ...959...39H}
}

@ARTICLE{Harikane+2025,
       author = {{Harikane}, Yuichi and {Inoue}, Akio K. and {Ellis}, Richard S. and {Ouchi}, Masami and {Nakazato}, Yurina and {Yoshida}, Naoki and {Ono}, Yoshiaki and {Sun}, Fengwu and {Sato}, Riku A. and {Ferrami}, Giovanni and {Fujimoto}, Seiji and {Kashikawa}, Nobunari and {McLeod}, Derek J. and {P{\'e}rez-Gonz{\'a}lez}, Pablo G. and {Sawicki}, Marcin and {Sugahara}, Yuma and {Xu}, Yi and {Yamanaka}, Satoshi and {Carnall}, Adam C. and {Cullen}, Fergus and {Dunlop}, James S. and {Egami}, Eiichi and {Grogin}, Norman and {Isobe}, Yuki and {Koekemoer}, Anton M. and {Laporte}, Nicolas and {Lee}, Chien-Hsiu and {Magee}, Dan and {Matsuo}, Hiroshi and {Matsuoka}, Yoshiki and {Mawatari}, Ken and {Nakajima}, Kimihiko and {Nakane}, Minami and {Tamura}, Yoichi and {Umeda}, Hiroya and {Yanagisawa}, Hiroto},
        title = "{JWST, ALMA, and Keck Spectroscopic Constraints on the UV Luminosity Functions at z {\ensuremath{\sim}} 7─14: Clumpiness and Compactness of the Brightest Galaxies in the Early Universe}",
      journal = {\apj},
         year = 2025,
        month = feb,
       volume = {980},
       number = {1},
          eid = {138},
        pages = {138},
          doi = {10.3847/1538-4357/ad9b2c},
archivePrefix = {arXiv},
       eprint = {2406.18352},
 primaryClass = {astro-ph.GA},
       adsurl = {https://ui.adsabs.harvard.edu/abs/2025ApJ...980..138H}
}

@ARTICLE{Stacy+2013,
       author = {{Stacy}, Athena and {Bromm}, Volker},
        title = "{Constraining the statistics of Population III binaries}",
      journal = {\mnras},
         year = 2013,
        month = aug,
       volume = {433},
       number = {2},
        pages = {1094-1107},
          doi = {10.1093/mnras/stt789},
archivePrefix = {arXiv},
       eprint = {1211.1889},
 primaryClass = {astro-ph.CO},
       adsurl = {https://ui.adsabs.harvard.edu/abs/2013MNRAS.433.1094S}
}

@ARTICLE{Klessen+2023,
       author = {{Klessen}, Ralf S. and {Glover}, Simon C.~O.},
        title = "{The First Stars: Formation, Properties, and Impact}",
      journal = {\araa},
         year = 2023,
        month = aug,
       volume = {61},
        pages = {65-130},
          doi = {10.1146/annurev-astro-071221-053453},
archivePrefix = {arXiv},
       eprint = {2303.12500},
 primaryClass = {astro-ph.CO},
       adsurl = {https://ui.adsabs.harvard.edu/abs/2023ARA&A..61...65K}
}

@ARTICLE{Nomoto+2013,
       author = {{Nomoto}, Ken'ichi and {Kobayashi}, Chiaki and {Tominaga}, Nozomu},
        title = "{Nucleosynthesis in Stars and the Chemical Enrichment of Galaxies}",
      journal = {\araa},
         year = 2013,
        month = aug,
       volume = {51},
       number = {1},
        pages = {457-509},
          doi = {10.1146/annurev-astro-082812-140956},
       adsurl = {https://ui.adsabs.harvard.edu/abs/2013ARA&A..51..457N}
}

@ARTICLE{Chiaki+2023,
       author = {{Chiaki}, Gen and {Chon}, Sunmyon and {Omukai}, Kazuyuki and {Trinca}, Alessandro and {Schneider}, Raffaella and {Valiante}, Rosa},
        title = "{Direct-collapse black hole formation induced by internal radiation of host haloes}",
      journal = {\mnras},
         year = 2023,
        month = may,
       volume = {521},
       number = {2},
        pages = {2845-2859},
          doi = {10.1093/mnras/stad689},
archivePrefix = {arXiv},
       eprint = {2303.01762},
 primaryClass = {astro-ph.GA},
       adsurl = {https://ui.adsabs.harvard.edu/abs/2023MNRAS.521.2845C}
}

@ARTICLE{Chon+2024,
       author = {{Chon}, Sunmyon and {Hosokawa}, Takashi and {Omukai}, Kazuyuki and {Schneider}, Raffaella},
        title = "{Impact of radiative feedback on the initial mass function of metal-poor stars}",
      journal = {\mnras},
         year = 2024,
        month = may,
       volume = {530},
       number = {3},
        pages = {2453-2474},
          doi = {10.1093/mnras/stae1027},
archivePrefix = {arXiv},
       eprint = {2312.13339},
 primaryClass = {astro-ph.GA},
       adsurl = {https://ui.adsabs.harvard.edu/abs/2024MNRAS.530.2453C}
}

@ARTICLE{Latif+2018,
       author = {{Latif}, Muhammad A. and {Volonteri}, Marta and {Wise}, John H.},
        title = "{Early growth of typical high-redshift black holes seeded by direct collapse}",
      journal = {\mnras},
         year = 2018,
        month = jun,
       volume = {476},
       number = {4},
        pages = {5016-5025},
          doi = {10.1093/mnras/sty622},
archivePrefix = {arXiv},
       eprint = {1801.07685},
 primaryClass = {astro-ph.GA},
       adsurl = {https://ui.adsabs.harvard.edu/abs/2018MNRAS.476.5016L}
}

@ARTICLE{Chon+2021,
       author = {{Chon}, Sunmyon and {Omukai}, Kazuyuki and {Schneider}, Raffaella},
        title = "{Transition of the initial mass function in the metal-poor environments}",
      journal = {\mnras},
         year = 2021,
        month = dec,
       volume = {508},
       number = {3},
        pages = {4175-4192},
          doi = {10.1093/mnras/stab2497},
archivePrefix = {arXiv},
       eprint = {2103.04997},
 primaryClass = {astro-ph.GA},
       adsurl = {https://ui.adsabs.harvard.edu/abs/2021MNRAS.508.4175C}
}

@ARTICLE{Chon+2021a,
       author = {{Chon}, Sunmyon and {Hosokawa}, Takashi and {Omukai}, Kazuyuki},
        title = "{Cosmological direct-collapse black hole formation sites hostile for their growth}",
      journal = {\mnras},
         year = 2021,
        month = mar,
       volume = {502},
       number = {1},
        pages = {700-713},
          doi = {10.1093/mnras/stab061},
archivePrefix = {arXiv},
       eprint = {2008.09120},
 primaryClass = {astro-ph.GA},
       adsurl = {https://ui.adsabs.harvard.edu/abs/2021MNRAS.502..700C}
}

@ARTICLE{Chon+2022,
       author = {{Chon}, Sunmyon and {Ono}, Haruka and {Omukai}, Kazuyuki and {Schneider}, Raffaella},
        title = "{Impact of the cosmic background radiation on the initial mass function of metal-poor stars}",
      journal = {\mnras},
         year = 2022,
        month = aug,
       volume = {514},
       number = {3},
        pages = {4639-4654},
          doi = {10.1093/mnras/stac1549},
archivePrefix = {arXiv},
       eprint = {2205.15328},
 primaryClass = {astro-ph.GA},
       adsurl = {https://ui.adsabs.harvard.edu/abs/2022MNRAS.514.4639C}
}

@ARTICLE{Bellovary+2010,
       author = {{Bellovary}, Jillian M. and {Governato}, Fabio and {Quinn}, Thomas R. and {Wadsley}, James and {Shen}, Sijing and {Volonteri}, Marta},
        title = "{Wandering Black Holes in Bright Disk Galaxy Halos}",
      journal = {\apjl},
         year = 2010,
        month = oct,
       volume = {721},
       number = {2},
        pages = {L148-L152},
          doi = {10.1088/2041-8205/721/2/L148},
archivePrefix = {arXiv},
       eprint = {1008.5147},
 primaryClass = {astro-ph.CO},
       adsurl = {https://ui.adsabs.harvard.edu/abs/2010ApJ...721L.148B}
}

@ARTICLE{Tremmel+2018,
       author = {{Tremmel}, Michael and {Governato}, Fabio and {Volonteri}, Marta and {Pontzen}, Andrew and {Quinn}, Thomas R.},
        title = "{Wandering Supermassive Black Holes in Milky-Way-mass Halos}",
      journal = {\apjl},
         year = 2018,
        month = apr,
       volume = {857},
       number = {2},
          eid = {L22},
        pages = {L22},
          doi = {10.3847/2041-8213/aabc0a},
archivePrefix = {arXiv},
       eprint = {1802.06783},
 primaryClass = {astro-ph.GA},
       adsurl = {https://ui.adsabs.harvard.edu/abs/2018ApJ...857L..22T}
}

@ARTICLE{Ricarte+2021,
       author = {{Ricarte}, Angelo and {Tremmel}, Michael and {Natarajan}, Priyamvada and {Zimmer}, Charlotte and {Quinn}, Thomas},
        title = "{Origins and demographics of wandering black holes}",
      journal = {\mnras},
         year = 2021,
        month = jun,
       volume = {503},
       number = {4},
        pages = {6098-6111},
          doi = {10.1093/mnras/stab866},
archivePrefix = {arXiv},
       eprint = {2103.12124},
 primaryClass = {astro-ph.GA},
       adsurl = {https://ui.adsabs.harvard.edu/abs/2021MNRAS.503.6098R}
}

@ARTICLE{Hosokawa+2013,
       author = {{Hosokawa}, Takashi and {Yorke}, Harold W. and {Inayoshi}, Kohei and {Omukai}, Kazuyuki and {Yoshida}, Naoki},
        title = "{Formation of Primordial Supermassive Stars by Rapid Mass Accretion}",
      journal = {\apj},
         year = 2013,
        month = dec,
       volume = {778},
       number = {2},
          eid = {178},
        pages = {178},
          doi = {10.1088/0004-637X/778/2/178},
archivePrefix = {arXiv},
       eprint = {1308.4457},
 primaryClass = {astro-ph.SR},
       adsurl = {https://ui.adsabs.harvard.edu/abs/2013ApJ...778..178H}
}

@ARTICLE{Fuller+1986,
       author = {{Fuller}, G.~M. and {Woosley}, S.~E. and {Weaver}, T.~A.},
        title = "{The Evolution of Radiation-dominated Stars. I. Nonrotating Supermassive Stars}",
      journal = {\apj},
         year = 1986,
        month = aug,
       volume = {307},
        pages = {675},
          doi = {10.1086/164452},
       adsurl = {https://ui.adsabs.harvard.edu/abs/1986ApJ...307..675F}
}

@ARTICLE{Shibata+2002,
       author = {{Shibata}, Masaru and {Shapiro}, Stuart L.},
        title = "{Collapse of a Rotating Supermassive Star to a Supermassive Black Hole: Fully Relativistic Simulations}",
      journal = {\apjl},
         year = 2002,
        month = jun,
       volume = {572},
       number = {1},
        pages = {L39-L43},
          doi = {10.1086/341516},
archivePrefix = {arXiv},
       eprint = {astro-ph/0205091},
 primaryClass = {astro-ph},
       adsurl = {https://ui.adsabs.harvard.edu/abs/2002ApJ...572L..39S}
}

@ARTICLE{Banados+2018,
       author = {{Ba{\~n}ados}, Eduardo and {Venemans}, Bram P. and {Mazzucchelli}, Chiara and {Farina}, Emanuele P. and {Walter}, Fabian and {Wang}, Feige and {Decarli}, Roberto and {Stern}, Daniel and {Fan}, Xiaohui and {Davies}, Frederick B. and {Hennawi}, Joseph F. and {Simcoe}, Robert A. and {Turner}, Monica L. and {Rix}, Hans-Walter and {Yang}, Jinyi and {Kelson}, Daniel D. and {Rudie}, Gwen C. and {Winters}, Jan Martin},
        title = "{An 800-million-solar-mass black hole in a significantly neutral Universe at a redshift of 7.5}",
      journal = {\nat},
         year = 2018,
        month = jan,
       volume = {553},
       number = {7689},
        pages = {473-476},
          doi = {10.1038/nature25180},
archivePrefix = {arXiv},
       eprint = {1712.01860},
 primaryClass = {astro-ph.GA},
       adsurl = {https://ui.adsabs.harvard.edu/abs/2018Natur.553..473B}
}

@ARTICLE{Wang+2021,
       author = {{Wang}, Feige and {Yang}, Jinyi and {Fan}, Xiaohui and {Hennawi}, Joseph F. and {Barth}, Aaron J. and {Banados}, Eduardo and {Bian}, Fuyan and {Boutsia}, Konstantina and {Connor}, Thomas and {Davies}, Frederick B. and {Decarli}, Roberto and {Eilers}, Anna-Christina and {Farina}, Emanuele Paolo and {Green}, Richard and {Jiang}, Linhua and {Li}, Jiang-Tao and {Mazzucchelli}, Chiara and {Nanni}, Riccardo and {Schindler}, Jan-Torge and {Venemans}, Bram and {Walter}, Fabian and {Wu}, Xue-Bing and {Yue}, Minghao},
        title = "{A Luminous Quasar at Redshift 7.642}",
      journal = {\apjl},
         year = 2021,
        month = jan,
       volume = {907},
       number = {1},
          eid = {L1},
        pages = {L1},
          doi = {10.3847/2041-8213/abd8c6},
archivePrefix = {arXiv},
       eprint = {2101.03179},
 primaryClass = {astro-ph.GA},
       adsurl = {https://ui.adsabs.harvard.edu/abs/2021ApJ...907L...1W}
}

@ARTICLE{Bogdan+2024,
       author = {{Bogd{\'a}n}, {\'A}kos and {Goulding}, Andy D. and {Natarajan}, Priyamvada and {Kov{\'a}cs}, Orsolya E. and {Tremblay}, Grant R. and {Chadayammuri}, Urmila and {Volonteri}, Marta and {Kraft}, Ralph P. and {Forman}, William R. and {Jones}, Christine and {Churazov}, Eugene and {Zhuravleva}, Irina},
        title = "{Evidence for heavy-seed origin of early supermassive black holes from a z ≍ 10 X-ray quasar}",
      journal = {Nature Astronomy},
         year = 2024,
        month = jan,
       volume = {8},
       number = {1},
        pages = {126-133},
          doi = {10.1038/s41550-023-02111-9},
archivePrefix = {arXiv},
       eprint = {2305.15458},
 primaryClass = {astro-ph.GA},
       adsurl = {https://ui.adsabs.harvard.edu/abs/2024NatAs...8..126B}
}

@ARTICLE{Mortlock+2011,
       author = {{Mortlock}, Daniel J. and {Warren}, Stephen J. and {Venemans}, Bram P. and {Patel}, Mitesh and {Hewett}, Paul C. and {McMahon}, Richard G. and {Simpson}, Chris and {Theuns}, Tom and {Gonz{\'a}les-Solares}, Eduardo A. and {Adamson}, Andy and {Dye}, Simon and {Hambly}, Nigel C. and {Hirst}, Paul and {Irwin}, Mike J. and {Kuiper}, Ernst and {Lawrence}, Andy and {R{\"o}ttgering}, Huub J.~A.},
        title = "{A luminous quasar at a redshift of z = 7.085}",
      journal = {\nat},
         year = 2011,
        month = jun,
       volume = {474},
       number = {7353},
        pages = {616-619},
          doi = {10.1038/nature10159},
archivePrefix = {arXiv},
       eprint = {1106.6088},
 primaryClass = {astro-ph.CO},
       adsurl = {https://ui.adsabs.harvard.edu/abs/2011Natur.474..616M}
}

@ARTICLE{Hosokawa+2012,
       author = {{Hosokawa}, Takashi and {Omukai}, Kazuyuki and {Yorke}, Harold W.},
        title = "{Rapidly Accreting Supergiant Protostars: Embryos of Supermassive Black Holes?}",
      journal = {\apj},
         year = 2012,
        month = sep,
       volume = {756},
       number = {1},
          eid = {93},
        pages = {93},
          doi = {10.1088/0004-637X/756/1/93},
archivePrefix = {arXiv},
       eprint = {1203.2613},
 primaryClass = {astro-ph.CO},
       adsurl = {https://ui.adsabs.harvard.edu/abs/2012ApJ...756...93H}
}

@ARTICLE{Chabrier2003,
       author = {{Chabrier}, Gilles},
        title = "{Galactic Stellar and Substellar Initial Mass Function}",
      journal = {\pasp},
         year = 2003,
        month = jul,
       volume = {115},
       number = {809},
        pages = {763-795},
          doi = {10.1086/376392},
archivePrefix = {arXiv},
       eprint = {astro-ph/0304382},
 primaryClass = {astro-ph},
       adsurl = {https://ui.adsabs.harvard.edu/abs/2003PASP..115..763C}
}

@ARTICLE{Chon+2020,
       author = {{Chon}, Sunmyon and {Omukai}, Kazuyuki},
        title = "{Supermassive star formation via super competitive accretion in slightly metal-enriched clouds}",
      journal = {\mnras},
         year = 2020,
        month = may,
       volume = {494},
       number = {2},
        pages = {2851-2860},
          doi = {10.1093/mnras/staa863},
archivePrefix = {arXiv},
       eprint = {2001.06491},
 primaryClass = {astro-ph.GA},
       adsurl = {https://ui.adsabs.harvard.edu/abs/2020MNRAS.494.2851C}
}

@ARTICLE{Chon+2025,
       author = {{Chon}, Sunmyon and {Omukai}, Kazuyuki},
        title = "{Formation of supermassive stars and dense star clusters in metal-poor clouds exposed to strong FUV radiation}",
      journal = {\mnras},
         year = 2025,
        month = may,
       volume = {539},
       number = {3},
        pages = {2561-2582},
          doi = {10.1093/mnras/staf598},
archivePrefix = {arXiv},
       eprint = {2412.14900},
 primaryClass = {astro-ph.GA},
       adsurl = {https://ui.adsabs.harvard.edu/abs/2025MNRAS.539.2561C}
}

@ARTICLE{Chon+2026,
       author = {{Chon}, Sunmyon and {Hirano}, Shingo and {Ishiyama}, Tomoaki and {Chang}, Seok-Jun and {Springel}, Volker},
        title = "{Rapid emergence of overmassive black holes in the early Universe}",
      journal = {arXiv e-prints},
         year = 2026,
        month = jan,
          eid = {arXiv:2601.04955},
        pages = {arXiv:2601.04955},
          doi = {10.48550/arXiv.2601.04955},
archivePrefix = {arXiv},
       eprint = {2601.04955},
 primaryClass = {astro-ph.GA},
       adsurl = {https://ui.adsabs.harvard.edu/abs/2026arXiv260104955C}
}

@ARTICLE{Taylor+2025,
       author = {{Taylor}, Anthony J. and {Finkelstein}, Steven L. and {Kocevski}, Dale D. and {Jeon}, Junehyoung and {Bromm}, Volker and {Amor{\'\i}n}, Ricardo O. and {Arrabal Haro}, Pablo and {Backhaus}, Bren E. and {Bagley}, Micaela B. and {Banados}, Eduardo and {Bhatawdekar}, Rachana and {Brooks}, Madisyn and {Calabr{\`o}}, Antonello and {Ch{\'a}vez Ortiz}, {\'O}scar A. and {Cheng}, Yingjie and {Cleri}, Nikko J. and {Cole}, Justin W. and {Davis}, Kelcey and {Dickinson}, Mark and {Donnan}, Callum and {Dunlop}, James S. and {Ellis}, Richard S. and {Fern{\'a}ndez}, Vital and {Fontana}, Adriano and {Fujimoto}, Seiji and {Giavalisco}, Mauro and {Grazian}, Andrea and {Guo}, Jingsong and {Hathi}, Nimish P. and {Holwerda}, Benne W. and {Hirschmann}, Michaela and {Inayoshi}, Kohei and {Kartaltepe}, Jeyhan S. and {Khusanova}, Yana and {Koekemoer}, Anton M. and {Kokorev}, Vasily and {Larson}, Rebecca L. and {Leung}, Gene C.~K. and {Lucas}, Ray A. and {McLeod}, Derek J. and {Napolitano}, Lorenzo and {Onoue}, Masafusa and {Pacucci}, Fabio and {Papovich}, Casey and {P{\'e}rez-Gonz{\'a}lez}, Pablo G. and {Pirzkal}, Nor and {Somerville}, Rachel S. and {Trump}, Jonathan R. and {Wilkins}, Stephen M. and {Yung}, L.~Y. Aaron and {Zhang}, Haowen},
        title = "{Broad-line AGNs at 3.5 < z < 6: The Black Hole Mass Function and a Connection with Little Red Dots}",
      journal = {\apj},
         year = 2025,
        month = jun,
       volume = {986},
       number = {2},
          eid = {165},
        pages = {165},
          doi = {10.3847/1538-4357/add15b},
archivePrefix = {arXiv},
       eprint = {2409.06772},
 primaryClass = {astro-ph.GA},
       adsurl = {https://ui.adsabs.harvard.edu/abs/2025ApJ...986..165T}
}

@ARTICLE{Yanagisawa+2026,
       author = {{Yanagisawa}, Hiroto and {Ouchi}, Masami and {Golubchik}, Miriam and {Oguri}, Masamune and {Fujimoto}, Seiji and {Kokorev}, Vasily and {Brammer}, Gabriel and {Sun}, Fengwu and {Nakane}, Minami and {Harikane}, Yuichi and {Umeda}, Hiroya and {Akins}, Hollis B. and {Atek}, Hakim and {Bauer}, Franz E. and {Brada{\v{c}}}, Maru{\v{s}}a and {Chisholm}, John and {Coe}, Dan and {Diego}, Jose M. and {Ferguson}, Henry C. and {Finkelstein}, Steven L. and {Furtak}, Lukas J. and {Inayoshi}, Kohei and {Koekemoer}, Anton M. and {Matthee}, Jorryt and {Naidu}, Rohan P. and {Ono}, Yoshiaki and {Pan}, Richard and {Richard}, Johan and {Robbins}, Luke and {Willott}, Chris and {Zitrin}, Adi and {Amor{\'\i}n}, Ricardo O. and {Bradley}, Larry D. and {Bromm}, Volker and {Conselice}, Christopher J. and {Dayal}, Pratika and {Kartaltepe}, Jeyhan S. and {Lopes}, Paulo A.~A. and {Lucas}, Ray A. and {Magdis}, Georgios E. and {Martis}, Nicholas S. and {Papovich}, Casey and {Schaerer}, Daniel and {Valentino}, Francesco and {Vanzella}, Eros and {Allingham}, Joseph F.~V. and {Grogin}, Norman A. and {Gonz{\'a}lez-Otero}, Mauro and {Ricotti}, Massimo and {Windhorst}, Rogier A.},
        title = "{VENUS: Two Faint Little Red Dots Separated by $\sim70\,\mathrm{pc}$ Hidden in a Single Lensed Galaxy at $z\sim7$}",
      journal = {arXiv e-prints},
         year = 2026,
        month = jan,
          eid = {arXiv:2601.06015},
        pages = {arXiv:2601.06015},
          doi = {10.48550/arXiv.2601.06015},
archivePrefix = {arXiv},
       eprint = {2601.06015},
 primaryClass = {astro-ph.GA},
       adsurl = {https://ui.adsabs.harvard.edu/abs/2026arXiv260106015Y}
}

@ARTICLE{Umeda+2026,
       author = {{Umeda}, Hiroya and {Inayoshi}, Kohei and {Harikane}, Yuichi and {Murase}, Kohta},
        title = "{A Black Hole Envelope Interpretation for Cosmological Demographics of Little Red Dots}",
      journal = {\apj},
         year = 2026,
        month = mar,
       volume = {999},
       number = {2},
          eid = {183},
        pages = {183},
          doi = {10.3847/1538-4357/ae4101},
archivePrefix = {arXiv},
       eprint = {2512.04208},
 primaryClass = {astro-ph.GA},
       adsurl = {https://ui.adsabs.harvard.edu/abs/2026ApJ...999..183U}
}

@ARTICLE{Adamo+2024,
       author = {{Adamo}, Angela and {Bradley}, Larry D. and {Vanzella}, Eros and {Claeyssens}, Ad{\'e}la{\"\i}de and {Welch}, Brian and {Diego}, Jose M. and {Mahler}, Guillaume and {Oguri}, Masamune and {Sharon}, Keren and {Abdurro'uf} and {Hsiao}, Tiger Yu-Yang and {Xu}, Xinfeng and {Messa}, Matteo and {Lassen}, Augusto E. and {Zackrisson}, Erik and {Brammer}, Gabriel and {Coe}, Dan and {Kokorev}, Vasily and {Ricotti}, Massimo and {Zitrin}, Adi and {Fujimoto}, Seiji and {Inoue}, Akio K. and {Resseguier}, Tom and {Rigby}, Jane R. and {Jim{\'e}nez-Teja}, Yolanda and {Windhorst}, Rogier A. and {Hashimoto}, Takuya and {Tamura}, Yoichi},
        title = "{Bound star clusters observed in a lensed galaxy 460 Myr after the Big Bang}",
      journal = {\nat},
         year = 2024,
        month = aug,
       volume = {632},
       number = {8025},
        pages = {513-516},
          doi = {10.1038/s41586-024-07703-7},
archivePrefix = {arXiv},
       eprint = {2401.03224},
 primaryClass = {astro-ph.GA},
       adsurl = {https://ui.adsabs.harvard.edu/abs/2024Natur.632..513A}
}

@ARTICLE{Morishita+2025,
       author = {{Morishita}, Takahiro and {Liu}, Zhaoran and {Stiavelli}, Massimo and {Treu}, Tommaso and {Bergamini}, Pietro and {Zhang}, Yechi},
        title = "{Pristine Massive Star Formation Caught at the Break of Cosmic Dawn}",
      journal = {arXiv e-prints},
         year = 2025,
        month = jul,
          eid = {arXiv:2507.10521},
        pages = {arXiv:2507.10521},
          doi = {10.48550/arXiv.2507.10521},
archivePrefix = {arXiv},
       eprint = {2507.10521},
 primaryClass = {astro-ph.CO},
       adsurl = {https://ui.adsabs.harvard.edu/abs/2025arXiv250710521M}
}

@ARTICLE{Nakajima+2026,
       author = {{Nakajima}, Kimihiko and {Ouchi}, Masami and {Harikane}, Yuichi and {Vanzella}, Eros and {Ono}, Yoshiaki and {Isobe}, Yuki and {Nishigaki}, Moka and {Tsujimoto}, Takuji and {Nakamura}, Fumitaka and {Xu}, Yi and {Umeda}, Hiroya and {Zhang}, Yechi},
        title = "{An ultra-faint, chemically primitive galaxy forming in the reionization era}",
      journal = {\nat},
         year = 2026,
        month = may,
       volume = {653},
       number = {8114},
        pages = {363-367},
          doi = {10.1038/s41586-026-10374-1},
archivePrefix = {arXiv},
       eprint = {2506.11846},
 primaryClass = {astro-ph.GA},
       adsurl = {https://ui.adsabs.harvard.edu/abs/2026Natur.653..363N}
}

@ARTICLE{Chemerynska+2024,
       author = {{Chemerynska}, Iryna and {Atek}, Hakim and {Dayal}, Pratika and {Furtak}, Lukas J. and {Feldmann}, Robert and {Greene}, Jenny E. and {Maseda}, Michael V. and {Nanayakkara}, Themiya and {Oesch}, Pascal A. and {Fujimoto}, Seiji and {Labb{\'e}}, Ivo and {Bezanson}, Rachel and {Brammer}, Gabriel and {Cutler}, Sam E. and {Leja}, Joel and {Pan}, Richard and {Price}, Sedona H. and {Wang}, Bingjie and {Weaver}, John R. and {Whitaker}, Katherine E.},
        title = "{The Extreme Low-mass End of the Mass─Metallicity Relation at z {\ensuremath{\sim}} 7}",
      journal = {\apjl},
         year = 2024,
        month = nov,
       volume = {976},
       number = {1},
          eid = {L15},
        pages = {L15},
          doi = {10.3847/2041-8213/ad8dc9},
archivePrefix = {arXiv},
       eprint = {2407.17110},
 primaryClass = {astro-ph.GA},
       adsurl = {https://ui.adsabs.harvard.edu/abs/2024ApJ...976L..15C}
}

@ARTICLE{Curti+2024,
       author = {{Curti}, Mirko and {Maiolino}, Roberto and {Curtis-Lake}, Emma and {Chevallard}, Jacopo and {Carniani}, Stefano and {D'Eugenio}, Francesco and {Looser}, Tobias J. and {Scholtz}, Jan and {Charlot}, Stephane and {Cameron}, Alex and {{\"U}bler}, Hannah and {Witstok}, Joris and {Boyett}, Kristian and {Laseter}, Isaac and {Sandles}, Lester and {Arribas}, Santiago and {Bunker}, Andrew and {Giardino}, Giovanna and {Maseda}, Michael V. and {Rawle}, Tim and {Rodr{\'\i}guez Del Pino}, Bruno and {Smit}, Renske and {Willott}, Chris J. and {Eisenstein}, Daniel J. and {Hausen}, Ryan and {Johnson}, Benjamin and {Rieke}, Marcia and {Robertson}, Brant and {Tacchella}, Sandro and {Williams}, Christina C. and {Willmer}, Christopher and {Baker}, William M. and {Bhatawdekar}, Rachana and {Egami}, Eiichi and {Helton}, Jakob M. and {Ji}, Zhiyuan and {Kumari}, Nimisha and {Perna}, Michele and {Shivaei}, Irene and {Sun}, Fengwu},
        title = "{JADES: Insights into the low-mass end of the mass-metallicity-SFR relation at 3 < z < 10 from deep JWST/NIRSpec spectroscopy}",
      journal = {\aap},
         year = 2024,
        month = apr,
       volume = {684},
          eid = {A75},
        pages = {A75},
          doi = {10.1051/0004-6361/202346698},
archivePrefix = {arXiv},
       eprint = {2304.08516},
 primaryClass = {astro-ph.GA},
       adsurl = {https://ui.adsabs.harvard.edu/abs/2024A&A...684A..75C}
}

@ARTICLE{Nandal&Loeb2026,
       author = {{Nandal}, Devesh and {Loeb}, Abraham},
        title = "{Supermassive Stars Match the Spectral Signatures of JWST's Little Red Dots}",
      journal = {\apj},
         year = 2026,
        month = feb,
       volume = {998},
       number = {1},
          eid = {124},
        pages = {124},
          doi = {10.3847/1538-4357/ae32f3},
archivePrefix = {arXiv},
       eprint = {2507.12618},
 primaryClass = {astro-ph.GA},
       adsurl = {https://ui.adsabs.harvard.edu/abs/2026ApJ...998..124N}
}

@ARTICLE{Fukushima+2020,
       author = {{Fukushima}, Hajime and {Hosokawa}, Takashi and {Chiaki}, Gen and {Omukai}, Kazuyuki and {Yoshida}, Naoki and {Kuiper}, Rolf},
        title = "{Formation of massive stars under protostellar radiation feedback: very metal-poor stars}",
      journal = {\mnras},
         year = 2020,
        month = sep,
       volume = {497},
       number = {1},
        pages = {829-845},
          doi = {10.1093/mnras/staa1994},
archivePrefix = {arXiv},
       eprint = {2004.02364},
 primaryClass = {astro-ph.GA},
       adsurl = {https://ui.adsabs.harvard.edu/abs/2020MNRAS.497..829F}
}

@ARTICLE{Fukushima+2021,
       author = {{Fukushima}, Hajime and {Yajima}, Hidenobu},
        title = "{Radiation hydrodynamics simulations of massive star cluster formation in giant molecular clouds}",
      journal = {\mnras},
         year = 2021,
        month = oct,
       volume = {506},
       number = {4},
        pages = {5512-5539},
          doi = {10.1093/mnras/stab2099},
archivePrefix = {arXiv},
       eprint = {2104.10892},
 primaryClass = {astro-ph.GA},
       adsurl = {https://ui.adsabs.harvard.edu/abs/2021MNRAS.506.5512F}
}

@ARTICLE{He+2019,
       author = {{He}, Chong-Chong and {Ricotti}, Massimo and {Geen}, Sam},
        title = "{Simulating star clusters across cosmic time - I. Initial mass function, star formation rates, and efficiencies}",
      journal = {\mnras},
         year = 2019,
        month = oct,
       volume = {489},
       number = {2},
        pages = {1880-1898},
          doi = {10.1093/mnras/stz2239},
archivePrefix = {arXiv},
       eprint = {1904.07889},
 primaryClass = {astro-ph.GA},
       adsurl = {https://ui.adsabs.harvard.edu/abs/2019MNRAS.489.1880H}
}

@ARTICLE{Fukushima+2023,
       author = {{Fukushima}, Hajime and {Yajima}, Hidenobu},
        title = "{The formation of globular clusters with top-heavy initial mass functions}",
      journal = {\mnras},
         year = 2023,
        month = sep,
       volume = {524},
       number = {1},
        pages = {1422-1430},
          doi = {10.1093/mnras/stad1956},
archivePrefix = {arXiv},
       eprint = {2303.12405},
 primaryClass = {astro-ph.GA},
       adsurl = {https://ui.adsabs.harvard.edu/abs/2023MNRAS.524.1422F}
}

@ARTICLE{Tanaka+2024,
       author = {{Tanaka}, Takumi S. and {Silverman}, John D. and {Shimasaku}, Kazuhiro and {Arita}, Junya and {Akins}, Hollis B. and {Wang}, Feige and {Inayoshi}, Kohei and {Ding}, Xuheng and {Onoue}, Masafusa and {Liu}, Zhaoxuan and {Casey}, Caitlin M. and {Lambrides}, Erini and {Kokorev}, Vasily and {Jin}, Shuowen and {Faisst}, Andreas L. and {Lyu}, Jianwei and {Schindler}, Jan-Torge and {Wu}, Yunjing and {Drakos}, Nicole and {Shen}, Yue and {Li}, Junyao and {Zhuang}, Mingyang and {Fei}, Qinyue and {Ito}, Kei and {Tee}, Wei Leong and {Liu}, Weizhe and {Ren}, Wenke and {Kiyota}, Tomokazu and {Li}, Zi-Jian and {Matsui}, Suin and {Ando}, Makoto and {Hatano}, Shun and {Fujii}, Michiko S. and {Kartaltepe}, Jeyhan S. and {Koekemoer}, Anton M. and {Liu}, Daizhong and {McCracken}, Henry Joy and {Rhodes}, Jason and {Robertson}, Brant E. and {Franco}, Maximilien and {Kakiichi}, Koki and {Yang}, Jinyi and {Meyer}, Romain A. and {Andika}, Irham T. and {Cloonan}, Aidan P. and {Fan}, Xiaohui and {Gozaliasl}, Ghassem and {Harish}, Santosh and {Hayward}, Christopher C. and {Huertas-Company}, Marc and {Kakkad}, Darshan and {Kinugawa}, Tomoya and {Li}, Mingyu and {Roy}, Namrata and {Shuntov}, Marko and {Talia}, Margherita and {Toft}, Sune and {Vijayan}, Aswin P. and {Zhang}, Yiyang},
        title = "{Hidden in Pixels I: Discovery of dual ``little red dots'' indicates excess clustering on kilo-parsec scales}",
      journal = {arXiv e-prints},
         year = 2024,
        month = dec,
          eid = {arXiv:2412.14246},
        pages = {arXiv:2412.14246},
          doi = {10.48550/arXiv.2412.14246},
archivePrefix = {arXiv},
       eprint = {2412.14246},
 primaryClass = {astro-ph.GA},
       adsurl = {https://ui.adsabs.harvard.edu/abs/2024arXiv241214246T}
}

@ARTICLE{Barger+2026,
       author = {{Barger}, A.~J. and {Cowie}, L.~L.},
        title = "{Double Dots: Compact Pairs Mark Little Red Dots and High-Redshift Broad-line AGNs}",
      journal = {arXiv e-prints},
         year = 2026,
        month = may,
          eid = {arXiv:2605.27903},
        pages = {arXiv:2605.27903},
          doi = {10.48550/arXiv.2605.27903},
archivePrefix = {arXiv},
       eprint = {2605.27903},
 primaryClass = {astro-ph.GA},
       adsurl = {https://ui.adsabs.harvard.edu/abs/2026arXiv260527903B}
}

@ARTICLE{Asada+2026,
       author = {{Asada}, Yoshihisa and {Inayoshi}, Kohei and {Fei}, Qinyue and {Fujimoto}, Seiji and {Willott}, Chris J.},
        title = "{Origins of the UV Continuum and Balmer Emission Lines in Little Red Dots: Observational Validation of Dense Gas Envelope Models Enshrouding the Active Galactic Nucleus}",
      journal = {\apjl},
         year = 2026,
        month = jul,
       volume = {1005},
       number = {1},
          eid = {L22},
        pages = {L22},
          doi = {10.3847/2041-8213/ae7d0a},
       adsurl = {https://ui.adsabs.harvard.edu/abs/2026ApJ..1005L..22A}
}

@ARTICLE{Menon+2024,
       author = {{Menon}, Shyam H. and {Lancaster}, Lachlan and {Burkhart}, Blakesley and {Somerville}, Rachel S. and {Dekel}, Avishai and {Krumholz}, Mark R.},
        title = "{The Interplay between the Initial Mass Function and Star Formation Efficiency through Radiative Feedback at High Stellar Surface Densities}",
      journal = {\apjl},
         year = 2024,
        month = jun,
       volume = {967},
       number = {2},
          eid = {L28},
        pages = {L28},
          doi = {10.3847/2041-8213/ad462d},
archivePrefix = {arXiv},
       eprint = {2405.00813},
 primaryClass = {astro-ph.GA},
       adsurl = {https://ui.adsabs.harvard.edu/abs/2024ApJ...967L..28M}
}

@ARTICLE{Hsiao+2025,
       author = {{Hsiao}, Tiger Yu-Yang and {Sun}, Fengwu and {Lin}, Xiaojing and {Coe}, Dan and {Egami}, Eiichi and {Eisenstein}, Daniel J. and {Fudamoto}, Yoshinobu and {Bunker}, Andrew J. and {Fan}, Xiaohui and {Harikane}, Yuichi and {Helton}, Jakob M. and {Kakiichi}, Koki and {Liu}, Yichen and {Liu}, Weizhe and {Maiolino}, Roberto and {Ouchi}, Masami and {Tee}, Wei Leong and {Wang}, Feige and {Wu}, Yunjing and {Xu}, Yi and {Yang}, Jinyi and {Zhu}, Yongda},
        title = "{SAPPHIRES: Extremely Metal-Poor Galaxy Candidates with $12+{\rm log(O/H)}<7.0$ at $z\sim5-7$ from Deep JWST/NIRCam Grism Observations}",
      journal = {arXiv e-prints},
         year = 2025,
        month = may,
          eid = {arXiv:2505.03873},
        pages = {arXiv:2505.03873},
          doi = {10.48550/arXiv.2505.03873},
archivePrefix = {arXiv},
       eprint = {2505.03873},
 primaryClass = {astro-ph.GA},
       adsurl = {https://ui.adsabs.harvard.edu/abs/2025arXiv250503873H}
}

@ARTICLE{Schaerer2002,
       author = {{Schaerer}, D.},
        title = "{On the properties of massive Population III stars and metal-free stellar populations}",
      journal = {\aap},
         year = 2002,
        month = jan,
       volume = {382},
        pages = {28-42},
          doi = {10.1051/0004-6361:20011619},
archivePrefix = {arXiv},
       eprint = {astro-ph/0110697},
 primaryClass = {astro-ph},
       adsurl = {https://ui.adsabs.harvard.edu/abs/2002A&A...382...28S}
}

@ARTICLE{Springel+2021,
       author = {{Springel}, Volker and {Pakmor}, R{\"u}diger and {Zier}, Oliver and {Reinecke}, Martin},
        title = "{Simulating cosmic structure formation with the GADGET-4 code}",
      journal = {\mnras},
         year = 2021,
        month = sep,
       volume = {506},
       number = {2},
        pages = {2871-2949},
          doi = {10.1093/mnras/stab1855},
archivePrefix = {arXiv},
       eprint = {2010.03567},
 primaryClass = {astro-ph.IM},
       adsurl = {https://ui.adsabs.harvard.edu/abs/2021MNRAS.506.2871S}
}

@ARTICLE{Cullen+2025,
       author = {{Cullen}, F. and {Carnall}, A.~C. and {Scholte}, D. and {McLeod}, D.~J. and {McLure}, R.~J. and {Arellano-C{\'o}rdova}, K.~Z. and {Stanton}, T.~M. and {Donnan}, C.~T. and {Dunlop}, J.~S. and {Shapley}, A.~E. and {Barrufet}, L. and {Begley}, R. and {Bondestam}, C. and {Cirasuolo}, M. and {Leung}, H.-H. and {Pollock}, C.~L. and {Stevenson}, S.},
        title = "{The JWST EXCELS survey: an extremely metal-poor galaxy at z = 8.271 hosting an unusual population of massive stars}",
      journal = {\mnras},
         year = 2025,
        month = jul,
       volume = {540},
       number = {3},
        pages = {2176-2194},
          doi = {10.1093/mnras/staf838},
archivePrefix = {arXiv},
       eprint = {2501.11099},
 primaryClass = {astro-ph.GA},
       adsurl = {https://ui.adsabs.harvard.edu/abs/2025MNRAS.540.2176C}
}

@ARTICLE{Vogelsberger+2014,
       author = {{Vogelsberger}, Mark and {Genel}, Shy and {Springel}, Volker and {Torrey}, Paul and {Sijacki}, Debora and {Xu}, Dandan and {Snyder}, Greg and {Nelson}, Dylan and {Hernquist}, Lars},
        title = "{Introducing the Illustris Project: simulating the coevolution of dark and visible matter in the Universe}",
      journal = {\mnras},
         year = 2014,
        month = oct,
       volume = {444},
       number = {2},
        pages = {1518-1547},
          doi = {10.1093/mnras/stu1536},
archivePrefix = {arXiv},
       eprint = {1405.2921},
 primaryClass = {astro-ph.CO},
       adsurl = {https://ui.adsabs.harvard.edu/abs/2014MNRAS.444.1518V}
}

@ARTICLE{Schaye+2015,
       author = {{Schaye}, Joop and {Crain}, Robert A. and {Bower}, Richard G. and {Furlong}, Michelle and {Schaller}, Matthieu and {Theuns}, Tom and {Dalla Vecchia}, Claudio and {Frenk}, Carlos S. and {McCarthy}, I.~G. and {Helly}, John C. and {Jenkins}, Adrian and {Rosas-Guevara}, Y.~M. and {White}, Simon D.~M. and {Baes}, Maarten and {Booth}, C.~M. and {Camps}, Peter and {Navarro}, Julio F. and {Qu}, Yan and {Rahmati}, Alireza and {Sawala}, Till and {Thomas}, Peter A. and {Trayford}, James},
        title = "{The EAGLE project: simulating the evolution and assembly of galaxies and their environments}",
      journal = {\mnras},
         year = 2015,
        month = jan,
       volume = {446},
       number = {1},
        pages = {521-554},
          doi = {10.1093/mnras/stu2058},
archivePrefix = {arXiv},
       eprint = {1407.7040},
 primaryClass = {astro-ph.GA},
       adsurl = {https://ui.adsabs.harvard.edu/abs/2015MNRAS.446..521S}
}

@ARTICLE{Weinberger+2017,
       author = {{Weinberger}, Rainer and {Springel}, Volker and {Hernquist}, Lars and {Pillepich}, Annalisa and {Marinacci}, Federico and {Pakmor}, R{\"u}diger and {Nelson}, Dylan and {Genel}, Shy and {Vogelsberger}, Mark and {Naiman}, Jill and {Torrey}, Paul},
        title = "{Simulating galaxy formation with black hole driven thermal and kinetic feedback}",
      journal = {\mnras},
         year = 2017,
        month = mar,
       volume = {465},
       number = {3},
        pages = {3291-3308},
          doi = {10.1093/mnras/stw2944},
archivePrefix = {arXiv},
       eprint = {1607.03486},
 primaryClass = {astro-ph.GA},
       adsurl = {https://ui.adsabs.harvard.edu/abs/2017MNRAS.465.3291W}
}

@ARTICLE{Prole+2026,
       author = {{Prole}, Lewis Robert and {Regan}, John A. and {Mehta}, Daxal and {Pakmor}, Rudiger and {Koudmani}, Sophie and {Bourne}, Martin A. and {Glover}, Simon C.~O. and {Wise}, John H. and {Klessen}, Ralf S. and {Tremmel}, Michael and {Sijacki}, Debora and {Beckmann}, Ricarda S. and {Haehnelt}, Martin G. and {Brennan}, John and {van de Bor}, Pelle and {Clark}, Paul C.},
        title = "{The SEEDZ Simulations: Methodology and First Results on Massive Black Hole Seeding and Early Galaxy Growth}",
      journal = {The Open Journal of Astrophysics},
         year = 2026,
        month = mar,
       volume = {9},
        pages = {59236},
          doi = {10.33232/001c.159236},
archivePrefix = {arXiv},
       eprint = {2511.09640},
 primaryClass = {astro-ph.GA},
       adsurl = {https://ui.adsabs.harvard.edu/abs/2026OJAp....959236P}
}

@ARTICLE{Prole+2026b,
       author = {{Prole}, Lewis R. and {Regan}, John A. and {Mehta}, Daxal and {Nandal}, Devesh and {Pakmor}, R{\"u}diger and {Beckmann}, Ricarda S. and {Tremmel}, Michael and {Haehnelt}, Martin G. and {Glover}, Simon C.~O. and {Klessen}, Ralf S. and {Wise}, John H. and {Koudmani}, Sophie and {Bourne}, Martin A. and {Sijacki}, Debora and {Brennan}, John and {van de Bor}, Pelle and {Clark}, Paul C.},
        title = "{SEEDZ: Rapid Galaxy Assembly as a Pathway to Supermassive Stars, Dense Stellar Environments and Massive Black Hole Seeds}",
      journal = {arXiv e-prints},
         year = 2026,
        month = jun,
          eid = {arXiv:2606.27427},
        pages = {arXiv:2606.27427},
          doi = {10.48550/arXiv.2606.27427},
archivePrefix = {arXiv},
       eprint = {2606.27427},
 primaryClass = {astro-ph.GA},
       adsurl = {https://ui.adsabs.harvard.edu/abs/2026arXiv260627427P}
}

@ARTICLE{Mowla+2024,
       author = {{Mowla}, Lamiya and {Iyer}, Kartheik and {Asada}, Yoshihisa and {Desprez}, Guillaume and {Tan}, Vivian Yun Yan and {Martis}, Nicholas and {Sarrouh}, Ghassan and {Strait}, Victoria and {Abraham}, Roberto and {Brada{\v{c}}}, Maru{\v{s}}a and {Brammer}, Gabriel and {Muzzin}, Adam and {Pacifici}, Camilla and {Ravindranath}, Swara and {Sawicki}, Marcin and {Willott}, Chris and {Estrada-Carpenter}, Vince and {Jahan}, Nusrath and {Noirot}, Ga{\"e}l and {Matharu}, Jasleen and {Rihtar{\v{s}}i{\v{c}}}, Gregor and {Zabl}, Johannes},
        title = "{Formation of a low-mass galaxy from star clusters in a 600-million-year-old Universe}",
      journal = {\nat},
         year = 2024,
        month = dec,
       volume = {636},
       number = {8042},
        pages = {332-336},
          doi = {10.1038/s41586-024-08293-0},
archivePrefix = {arXiv},
       eprint = {2402.08696},
 primaryClass = {astro-ph.GA},
       adsurl = {https://ui.adsabs.harvard.edu/abs/2024Natur.636..332M}
}

@ARTICLE{Willott+2025,
       author = {{Willott}, Chris J. and {Asada}, Yoshihisa and {Iyer}, Kartheik G. and {Jude{\v{z}}}, Jon and {Rihtar{\v{s}}i{\v{c}}}, Gregor and {Martis}, Nicholas S. and {Sarrouh}, Ghassan T.~E. and {Desprez}, Guillaume and {Harshan}, Anishya and {Mowla}, Lamiya and {Noirot}, Ga{\"e}l and {Felicioni}, Giordano and {Brada{\v{c}}}, Maru{\v{s}}a and {Brammer}, Gabe and {Muzzin}, Adam and {Sawicki}, Marcin and {Antwi-Danso}, Jacqueline and {Markov}, Vladan and {Tripodi}, Roberta},
        title = "{In Search of the First Stars: An Ultra-compact and Very-low-metallicity Ly{\ensuremath{\alpha}} Emitter Deep within the Epoch of Reionization}",
      journal = {\apj},
         year = 2025,
        month = jul,
       volume = {988},
       number = {1},
          eid = {26},
        pages = {26},
          doi = {10.3847/1538-4357/addf49},
archivePrefix = {arXiv},
       eprint = {2502.07733},
 primaryClass = {astro-ph.GA},
       adsurl = {https://ui.adsabs.harvard.edu/abs/2025ApJ...988...26W}
}

@ARTICLE{Vanzella+2026,
       author = {{Vanzella}, E. and {Messa}, M. and {Zanella}, A. and {Bolamperti}, A. and {Castellano}, M. and {Loiacono}, F. and {Bergamini}, P. and {Roberts Borsani}, G. and {Adamo}, A. and {Fontana}, A. and {Treu}, T. and {Calura}, F. and {Grillo}, C. and {Lombardi}, M. and {Rosati}, P. and {Gilli}, R. and {Meneghetti}, M.},
        title = "{A pristine, star-forming complex at z = 4.19}",
      journal = {\aap},
         year = 2026,
        month = jan,
       volume = {705},
          eid = {L12},
        pages = {L12},
          doi = {10.1051/0004-6361/202557153},
archivePrefix = {arXiv},
       eprint = {2509.07073},
 primaryClass = {astro-ph.GA},
       adsurl = {https://ui.adsabs.harvard.edu/abs/2026A&A...705L..12V}
}

@ARTICLE{Cai+2025,
       author = {{Cai}, Sijia and {Li}, Mingyu and {Cai}, Zheng and {Wu}, Yunjing and {Yu}, Fujiang and {Dickinson}, Mark and {Sun}, Fengwu and {Fan}, Xiaohui and {Wang}, Ben and {Cullen}, Fergus and {Bian}, Fuyan and {Lin}, Xiaojing and {Zou}, Jiaqi},
        title = "{A Metal-free Galaxy at z = 3.19? Evidence of Late Population III Star Formation at Cosmic Noon}",
      journal = {\apjl},
         year = 2025,
        month = nov,
       volume = {993},
       number = {2},
          eid = {L52},
        pages = {L52},
          doi = {10.3847/2041-8213/ae1608},
archivePrefix = {arXiv},
       eprint = {2507.17820},
 primaryClass = {astro-ph.GA},
       adsurl = {https://ui.adsabs.harvard.edu/abs/2025ApJ...993L..52C}
}

@ARTICLE{Trussler+2026,
       author = {{Trussler}, James A.~A. and {Cameron}, Alex J. and {Eisenstein}, Daniel J. and {Katz}, Harley and {Adams}, Nathan J. and {Austin}, Duncan and {Bunker}, Andrew J. and {Carniani}, Stefano and {Conselice}, Christopher J. and {Curti}, Mirko and {Curtis-Lake}, Emma and {Hainline}, Kevin and {Harvey}, Thomas and {Johnson}, Benjamin D. and {Li}, Qiong and {Looser}, Tobias J. and {Rinaldi}, Pierluigi and {Robertson}, Brant and {Sun}, Fengwu and {Tacchella}, Sandro and {Williams}, Christina C. and {Willmer}, Christopher N.~A. and {Willott}, Chris and {Wu}, Zihao},
        title = "{Possible photometric signatures of nebular-dominated emission in 1.5 < z < 8.5 JADES galaxies}",
      journal = {\mnras},
         year = 2026,
        month = jun,
       volume = {549},
       number = {1},
          eid = {stag788},
        pages = {stag788},
          doi = {10.1093/mnras/stag788},
archivePrefix = {arXiv},
       eprint = {2510.12622},
 primaryClass = {astro-ph.GA},
       adsurl = {https://ui.adsabs.harvard.edu/abs/2026MNRAS.549ag788T}
}

@ARTICLE{Vanzella+2022,
       author = {{Vanzella}, E. and {Castellano}, M. and {Bergamini}, P. and {Meneghetti}, M. and {Zanella}, A. and {Calura}, F. and {Caminha}, G.~B. and {Rosati}, P. and {Cupani}, G. and {Me{\v{s}}tri{\'c}}, U. and {Brammer}, G. and {Tozzi}, P. and {Mercurio}, A. and {Grillo}, C. and {Sani}, E. and {Cristiani}, S. and {Nonino}, M. and {Merlin}, E. and {Pignataro}, G.~V.},
        title = "{High star cluster formation efficiency in the strongly lensed Sunburst Lyman-continuum galaxy at z = 2.37}",
      journal = {\aap},
         year = 2022,
        month = mar,
       volume = {659},
          eid = {A2},
        pages = {A2},
          doi = {10.1051/0004-6361/202141590},
archivePrefix = {arXiv},
       eprint = {2106.10280},
 primaryClass = {astro-ph.GA},
       adsurl = {https://ui.adsabs.harvard.edu/abs/2022A&A...659A...2V}
}

@ARTICLE{Vanzella+2023,
       author = {{Vanzella}, Eros and {Claeyssens}, Ad{\'e}la{\"\i}de and {Welch}, Brian and {Adamo}, Angela and {Coe}, Dan and {Diego}, Jose M. and {Mahler}, Guillaume and {Khullar}, Gourav and {Kokorev}, Vasily and {Oguri}, Masamune and {Ravindranath}, Swara and {Furtak}, Lukas J. and {Hsiao}, Tiger Yu-Yang and {Abdurro'uf} and {Mandelker}, Nir and {Brammer}, Gabriel and {Bradley}, Larry D. and {Brada{\v{c}}}, Maru{\v{s}}a and {Conselice}, Christopher J. and {Dayal}, Pratika and {Nonino}, Mario and {Andrade-Santos}, Felipe and {Windhorst}, Rogier A. and {Pirzkal}, Nor and {Sharon}, Keren and {de Mink}, S.~E. and {Fujimoto}, Seiji and {Zitrin}, Adi and {Eldridge}, Jan J. and {Norman}, Colin},
        title = "{JWST/NIRCam Probes Young Star Clusters in the Reionization Era Sunrise Arc}",
      journal = {\apj},
         year = 2023,
        month = mar,
       volume = {945},
       number = {1},
          eid = {53},
        pages = {53},
          doi = {10.3847/1538-4357/acb59a},
archivePrefix = {arXiv},
       eprint = {2211.09839},
 primaryClass = {astro-ph.GA},
       adsurl = {https://ui.adsabs.harvard.edu/abs/2023ApJ...945...53V}
}
\bibliographystyle{aasjournalv7}




\end{document}